\documentclass[a4paper,11pt,oneside]{article}

\usepackage{proofread}
\usepackage{amsmath,amssymb,hyperref}
\usepackage{fullpage}
\usepackage{float}

\usepackage{amsmath,amssymb,hyperref}
\usepackage{subcaption}
\usepackage[normalem]{ulem}

\usepackage{fancyvrb}%https://tex.stackexchange.com/questions/122195/how-to-center-verbatim

\usepackage[colorinlistoftodos,textsize=footnotesize,textwidth=1.5\marginparwidth]{todonotes}

\usepackage{enotez} %end notes
\usepackage{float}

\usepackage[T1]{fontenc}
\usepackage{xcolor}
\usepackage{lmodern}
\usepackage{authblk}

\usepackage{todonotes}

\title{A discrete model for layered growth}
\author[1]{Davide Renzi}
\author[1]{Sonia Marfia}
\author[2]{Giuseppe Tomassetti}
\author[3]{Giuseppe Zurlo}
\affil[1]{Dipartimento di Ingegneria Civile, Informatica e delle Tecnologie Aeronautiche, Roma Tre University, Via Vito Volterra 62, Roma, 00146, Italy}
\affil[2]{Dipartimento di Ingegneria Industriale, Elettronica e Meccanica, Roma Tre University, Via Vito Volterra 62, Roma, 00146, Italy}
\affil[3]{School of Mathematical and Statistical Sciences, University of Galway, University Road, Galway, Ireland}
\begin{document}
\maketitle
\begin{abstract}

In this work we present a discrete model that captures the fundamental properties of additively manufactured solids in a minimal setting. The model is based on simplified kinematics and allows for the onset of incompatible deformations between discrete layers of an additively manufactured stack. Thanks to the discrete nature of the model, we obtain an averaged formulation of mechanical equilibrium for the growing stack, leading to closed-form solutions that are both analytically simple and physically transparent. In particular, we are able to explain the origin of residual stresses by the accumulation of incompatible deformations between adjacent layers. At the same time, we are able to formulate the technologically relevant inverse problem that provides the deposition protocol required to produce a desired state of internal stress in the manufactured stack. Another important aspect analyzed in the work is the role played by an ideal ``glue'' between the layers, whose presence is fundamental to prevent their sliding and whose mechanical behavior can quantitatively influence the final stress distribution in the stack. Although the model is an elementary approximation of additive manufacturing, its simplicity makes it possible to highlight how the controls exerted during deposition will have qualitative or quantitative effects on the final stress state of the stack. This understanding is crucial in shedding light on the complex mechanical behavior of additive manufactured solids. 

\end{abstract}

\section{Introduction}

The modeling of growth phenomena has been a scientific challenge for many years and continues to offer opportunities for development in various fields, such as biomechanics, construction science, and materials science. Growth refers to the change in mass of a body over time, and it is observed in many contexts. In biology, growth is fundamental to describing many aspects of life, from cell division to tumor growth in living beings. However, growth processes also appear in many physical phenomena, where new material is added to an evolving system, such as 3D printing, or where growth is associated with phase transition phenomena, such as crystal growth \cite{Amar1986, Kessler1988,Langer1980}. 

Growth phenomena can be classified into two categories: volumetric growth \cite{Taber1995, Humphrey2003, Cowin2004, Rodriguez1994, Epstein2000, Garikipati2004, DiCarlo2002} and surface growth \cite{Skalak1997, Bacigalupo2012, Ciarletta2013, Tomassetti2016, Abeyaratne2022, Zurlo2017, Zurlo2018, Truskinovsky2019}. In the former case, the addition or removal of matter takes places inside the bulk; in the latter, material is added or removed at the boundary of the body. 

Both volumetric and surface growth can result in residual stress. This is not always an undesired outcome. In may biological contexts, residual stress plays an important role: it helps maintaining the structural integrity and the mechanical stability of a tissue; in blood vessels, residual stress contributes to the maintenance of an optimal vessel diameter; residual stress can influence how tissues respond to injury and their subsequent healing. In many technological contexts residual stress is an important target, like for example in the process of hot shrinking of metals, in reinforcing concrete structures, or in prestressing glass layers to enhance their resistance to fracture. 

Apart of these (nowadays classical) applications that regard residual stress as a beneficial feature, the last decade has seen a blooming of technologies aimed at embedding target residual stress patterns into ``shape lifting'' structures, an exciting avenue with an immense scenario of futuristic applications, such as 4D printing technologies \color{red}\cite{ge2016}, \color{black}biomimetic tissues and more.

The mathematical description of residual stress relies ultimately on  the concept of strain incompatibility, a measure of the level of ``unfitness'' of sub-parts of the body to be arranged together without creating voids or overlaps. In principle, this concept is purely geometrical, and then depending on the stiffness of the various parts of the body, the same level of incompatibility will ultimately result into different distributions of residual stress. 

Since the boundary of the body is directly accessible, it is intuitive that surface growth offers more control over the accretion process, giving the possibility of fine tuning, and hence program in advance the properties of the resulting body. Illustrating examples of how the choice of the accretion protocol affects the residual stresses may be found in \cite{Bacigalupo2012} in the context of 
masonry materials, and in \cite{Zurlo2017} in the context of 3D printing.

The understanding of the mechanism behind the accumulation of residual stress during surface growth was slowly achieved in stages over the span of the last 150 years. As early as 1883, G.H.Darwin (Charles' son) acknowledged the relevance of what he called ``the historical element'' in the horizontal thrust of a mass of sand \cite{darwinHorizontalThrustMass1883}, accounting for the fact that the final distribution of stress would depend on the way the sand had been layered in a container. 

The first analytical treatment of the deposition-dependent stress distribution in massive bodies is apparently traceable to E.I.Rashba \cite{Rashba1953}, 
who pioneered a prolific area of study of the residual stresses arising in surface growth problems. Some years later, Brown and Goodman \cite{Brown1963} recognize that the stress state of a massive, self-gravitating body formed by layered accretion is not the same as it would be if the body had been first fabricated, and then endowed with mass. This is due to the fact that when a new layer is deposited, it deforms the pre-existing material before hardening.  

The most recent developments in this vein have elucidated the geometric aspects of ``non-Euclidean'' surface growth \cite{Sozio2019,Sozio2020} and have established the non-locality, both in space and time, between the controls that are exerted on the growth surface and the ensuing state of incompatibility that is nailed in the body at the end of the process \cite{Zurlo2017}. What emerges from these studies is that strain incompatibility can not be controlled locally, in the sense that even assuming that the ``gluing'' takes place immediately after deposition, the local value of strain incompatibility depends not only on the pre-strain imposed on the new block prior to deposition, but also on the elastic adjustment of the underlying body, which is required to reattain equilibrium, and thus implicitly, also on the whole accretion history.

Three-dimensional problem involving incompatibility are often challenging to solve analytically, unless symmetry assumptions are made \cite{Dafalias2008,Dafalias2009,Tomassetti2016,Sozio2017}. However, one should be careful when making such assumptions, as symmetric solutions may be unstable during growth \cite{Abeyaratne2022a,Abeyaratne2022b}. One might consider simplifying the problem by reducing the number of dimensions to one. However, doing so would eliminate the issue of incompatibility altogether, which is of course not an option. As a solution \cite{Zurlo2018}, suggests retaining a trace of higher-dimensionality in the problem, through a 1.5-dimensional model. 

In the spirit of \cite{Zurlo2018} in the present paper we propose an approach where the growth of a body \emph{occurs primarily in one dimension} but the impact of the other dimensions is considered through the integration of \emph{extra state variables that account for stress and strain in the perpendicular directions}. Differently from previous studies, the problem is discrete \emph{both in space and time}. Specifically, the growing body is a stack elastic blocks glued sequentially, at discrete times, on top of another. It is quite intuitive that \color{black}when two elastic blocks are pre-strained, and then glued, residual stresses may arise depending on the loads that were applied to the blocks prior to attachment, and on the type of glue that was used in the process. A fast glue would freeze the state of the bonding interface at the very instant of adhesion, some other glues may allow for a partial or a total relaxation of stress before activating the bond. These intuitive considerations disclose the possibility to play with the order and manner in which an elastic body are assembled by the juxtaposition of elementary ``blocks'', in order to create desired distributions of residual stress in the resulting patch-worked body. This is the main idea explored in this paper.

We introduce our setup in Section 2, where we formulate the equilibrium problem for a weightless stack of $j$ blocks. 
Our key modeling assumptions are that each block undergoes a shear-less homogeneous strain, and that the difference between the horizontal strains of block $j+1$ and block $j$ be a prescribed \emph{incompatibility} $\delta_j$. Since the average  horizontal stress in each block vanishes 
the problem consists of oneglobal equilibrium condition, $j-1$ compatibility equations, and $j$ constitutive equations, for a total of $2j$ unknowns representing the horizontal stress and horizontal strain of the blocks. Of course this way to account for equilibrium in the horizontal direction is a crude approximation. In a real system it is inhomogeneous strain will occur, and therefore it should be understood that our analysis is only a first order approximation of a more refined model, to be developed elsewhere.

For this problem, we work out an analytical solution which shows that in the absence of external forces the horizontal stresses vanish only in the special case when the incompatibility vanishes as well. We then take into consideration external forces by extending the formulation to the case when each block is topped by a lumped mass, $b_i$. In this case, the vertical stresses must be taken into account, and can be computed by means of $j$ equilibrium conditions.

In Section 3, we shift our focus to the most fascinating issues in surface growth, specifically those where the incompatibilities  $\delta_i$, are not predefined, but instead dictated by the accretion rule. Starting from a stack of $j$ weighting blocks in equilibrium with prescribed body forces $b_i$, $i=1,\ldots,j$ and  incompatibilities $\delta_i$, $i=1,\ldots,j-1$, we position a new pre-stressed block atop the stack with a layer of glue between the two, and we subsequently release the block letting the whole system attain equilibrium.  The resulting stack will have $j+1$ block, where the incompatibility $\delta_j$ between block $j$ and block $j+1$ is determined by the character of the glue used for bonding. We consider two extreme cases: a ``fast'' glue and a ``slow'' glue.

We remark that the new block perturbs the equilibrium state of the underlying stack only \emph{after}  has been released. If the glue is fast, it locks the incompatibility $\delta_j$ \emph{before} the new block is released, when the strain of the new block is determined by its pre-stress, and the strain of the underlying stack is the same as in the previous accretion step. If the glue is very slow, on the other hand, the incompatibility $\delta_j$ locks in on only \emph{after} the block has been released. We assume that in the time span between the release of the new block and the solidification of the glue, the new block and the stack interact with each other with a \emph{frictionless} contact. As a result, when equilibrium is attained, the strain of the new block is determined by its own weight and not by its pre-stressed state; likewise, the strain of the underlying stack will change due to the additional weight of the new element. The strain difference between the last two blocks of the new stack determines the incompatibility after the glue solidifies.

Concerning the two modes of accretion, we work out explicit formulas for the incompatibility $\delta_j$. An interesting finding is that for the fast glue the incompatibility $\delta_j$ is determined by the pre-strain (and hence the pre-stress) in block $j+1$ and block $j$, as well as on the weights of the blocks. For the case of slow glue, instead, the incompatibility $\delta_j$ is determined solely by the weight $b_j$ of the $j-th$ element, and is unaffected, as anticipated, by the pre-stress.

Having established these general results, we consider in Section 4 some specific examples. In the first example we consider the growth of a stack up to the addition of 20 blocks, with constant horizontal tensile pre-stress, and we observe how the stress in the first block evolves as news blocks are added. What we observe is that the tensile pre-stress of the newly added blocks produces a contraction in the blocks beneath, and in particular, in the first block. However, as more and more blocks are added, the weight of the entire stack on the first block compensates the above-mentioned horizontal contraction, and as a result the horizontal strain of the first block becomes eventually positive. We also determine the strain profile in the entire stack, which shows that, while a constant incompatibility would result in a linear horizontal stress, a constant pre-stress produces a horizontal stress profile of ``parabolic'' type. In the second part of Section 4 we repeat the calculations for the case of slow glue, and we verify that the ``slow glue'' protocol results into lower final stressed.

We finally analyze the residual stresses that result from the removal of the weights, considering accretion protocols that differ not only for the type of glue used, but also for the weights of the blocks. If the weights $b_i$ are increasing, the profiles resulting from the ``fast'' and ``slow'' glue protocols are similar in shape, but differ by approximately an order of magnitude, with the residual stress being much larger in the case of ``fast glue''. If the weights are decreasing, on the other hand, the two profile have opposite convexities. Finally, if all weights are the same, the residual stress in the ``slow glue'' case is linear (since in this case the generated incompatibilities $\delta_i$ are all equal).

Section 5 deals with an even more intriguing problem: choosing a deposition strategy in order to attain a target residual stress, the so-called ``inverse problem''. As a first test, we start from the residual stress resulting from an analysis performed in Section 4 on the case when all weights are equal, and the pre-stress vanishes. We indeed verify that the original pre-stresses are recovered. As a further test, we consider the case of a sinusoidal residual stress, and we determine the pre-strains needed to obtain it. We then extend the model to incorporate Tresca's resistance criterion. We then repeat the treatment for the case of slow glue. Additional considerations and insights are contained in the concluding section.

\newpage

\section{Equilibrium of an incompatible stack}
\subsection{The notion of incompatibility}
\begin{figure}[thb!]
\centering
  \includegraphics[width=0.8\linewidth]{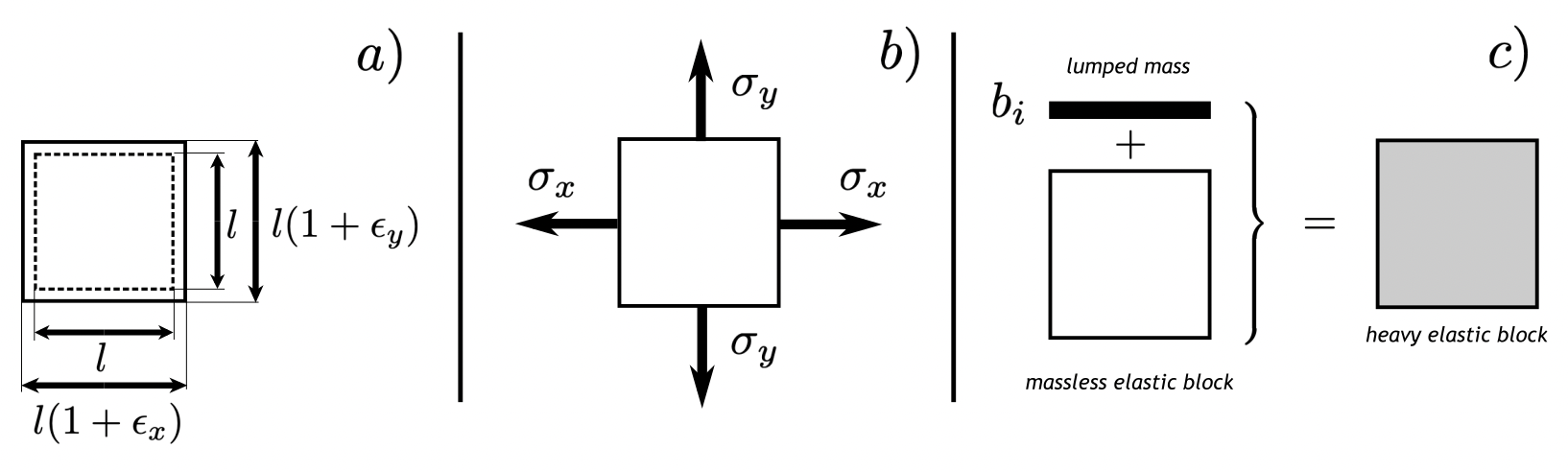} 
  \caption{\label{Fig1} \footnotesize{$a)$ reference (dashed) and strained (solid) block. $b)$ the block is assumed to deform homogeneously under the action of horizontal and vertical stresses $\sigma_x,\sigma_y$. $c)$ to simplify the notation, a grey block will denote the union of a lumped mass (of weight $b_i$) on top of a massless elastic block.}}
\end{figure}

%The main purpose of this section is to explain our notion of incompatibility. 
To set the stage, we consider a planar system consisting of a stack of blocks of linear elastic material,  which deform homogeneously and without shear strains.

We denote by $\epsilon_x^i$ and $\epsilon_y^i$ the horizontal and vertical strain of the $i$-th block. The corresponding stresses on the horizontal and vertical directions are $ \sigma_x^i$ and $ \sigma_y^i$, as shown in Fig.\ref{Fig1}.  The ``elastic state'' of the system is described by $2n$ \emph{strain components} and $2n$ \emph{stress components} , respectively,
%$$
%\{\epsilon_{x/y}^i,i=1,\ldots n\},
%$$ \normalsize
%$$
%\{\sigma_{x/y}^i,i=1,\ldots,n\}.
%$$
\begin{equation}
\mathcal E=\{(\epsilon_{x}^i,\epsilon_{y}^i),i=1,\ldots n\},\quad\text{and}\quad
\Sigma=\{(\sigma_{x}^i,\sigma_y^i),i=1,\ldots,n\}.
\end{equation}
We assume that the stress and the strain components are related by the \emph{costitutive equations}:
\begin{equation}
  \label{1}
 \begin{aligned}
   \epsilon_{x}^{i}=(\sigma_{x}^{i}-\nu\sigma_{y}^{i})/E\\
      \epsilon_{y}^{i}=(\sigma_{y}^{i}-\nu\sigma_{x}^{i})/E
   \end{aligned} \qquad i=1,...,n,
\end{equation}
where $E$ is the 2D \emph{Young modulus} (force/length) and $\nu$ is the \emph{Poisson ratio} (dimensionless). As a key ingredient in our model, we also assume that the stack comes with \emph{prescribed incompatibility} 
\begin{equation}
\Delta=\{\delta_i,i=1,\ldots,n-1\},	
\end{equation}
and we impose the \emph{incompatibility constraints}:
\begin{equation}\label{eq:1}
   \epsilon_x^{i+1}-\epsilon_x^i=\delta_i \qquad i=1,...,n-1,
  \end{equation}
These incompatibilities are maintained in the by a ``glue'', whose nature will be detailed in the sequel.
We shall see that, consistent with physical intuition, the enforcement of a non-trivial incompatibility $\Delta\neq 0$ induces a non-homogeneous strain state $\mathcal E$, as sketched in Fig. 2 below, and, as a result, a non-vanishing residual stress $\Sigma$. 
\begin{figure}[H]
    \centering
    \includegraphics[width=0.35\linewidth] {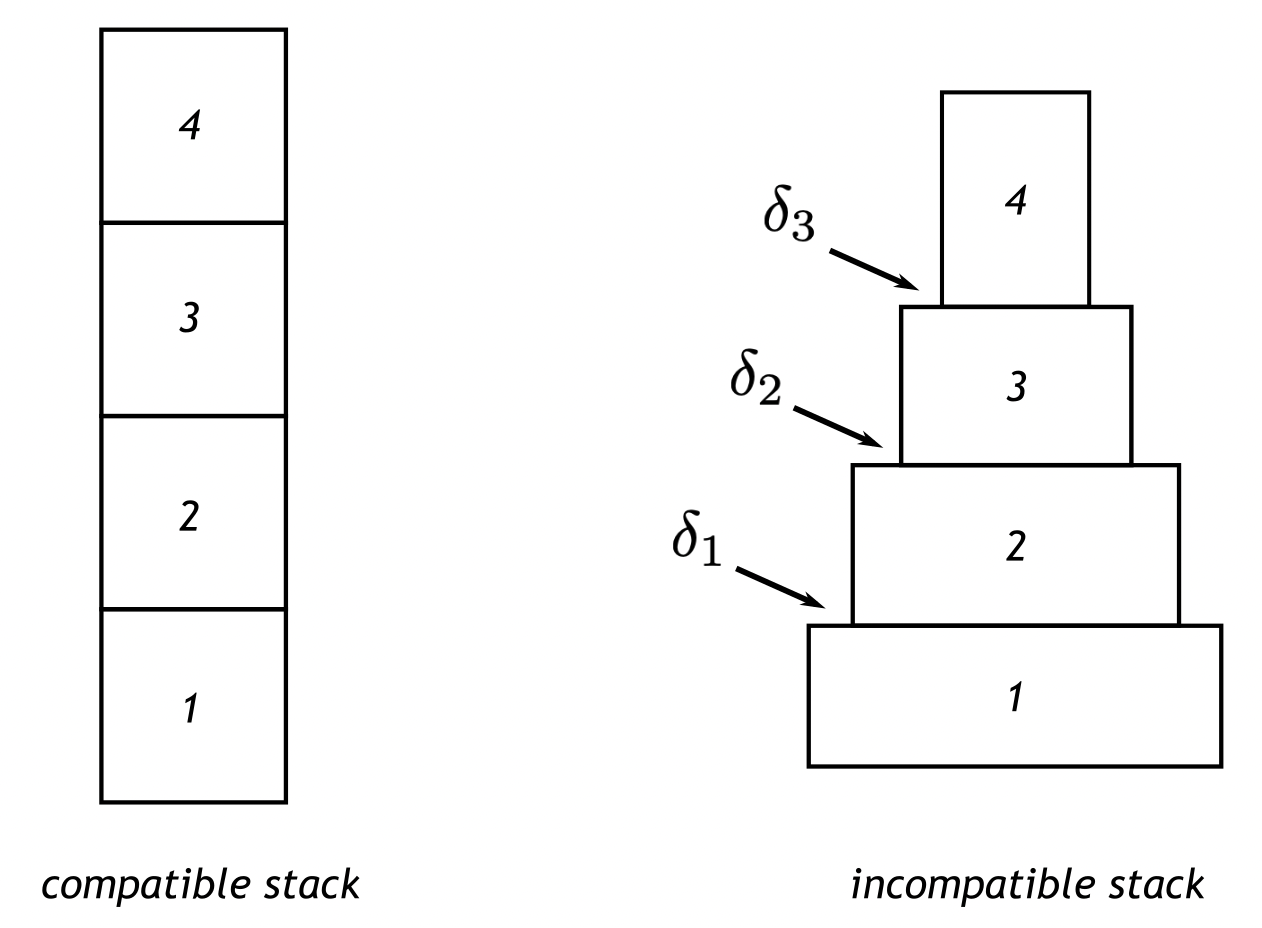}
    \caption{\label{Fig2}\footnotesize{Comparison between a compatible weightless stack, and an incompatible weightless stack with mismatches $\delta_i$ between the blocks $i$ and $i+1$}.}
\end{figure}

In this work we will focus attention both on weightless blocks, that according to Fig.\ref{Fig1}$_{a,b}$ will be described by unfilled blocks, and on heavy blocks, that will be filled in grey instead. In order to simplify the description of the state of deformation of the blocks under gravity, each block will be conceived as if it is formed by a heavy {\it lumped mass}, denoted in black in Fig.\ref{Fig1}$_c$, topping a massless block through a frictionless interface. This way we will be legitimated to keep describing the deformation of each block as homogeneous, whereas clearly in reality a heavy block with distributed mass will deform inhomogeneously.

The constitutive equations \eqref{1} and the incompatibility constraints \eqref{eq:1} are complemented by equilibrium equations, that we introduce in the next Section.

 \subsection{Weightless blocks}
We first consider the problem of equilibrium of a weightless stack, as the one represented in Fig.\ref{Fig2}. If the incompatibilities are prescribed and there are no external loads acting on the stack, to determine the state of stress we complement \eqref{eq:1} with the following $n+1$ equilibrium equations:
 \begin{equation}
\label{2}
\left\{
    \begin{array}{lll}
    {
    \displaystyle\sum_{i=1}^{{n}}\sigma_x^i=0},\\[1em]
    {\sigma_y^i=0 \qquad i=1,...,{n}}.
      \end{array}
\right.
\end{equation}
The first of \eqref{2} tells us that the stress average for the entire stack vanishes along the horizontal direction, whereas the remaining equations impose stress equilibrium of the individual blocks along the vertical direction. The averaged formulation of equilibrium \eqref{2} in the context of accreting media appeared for the first time in a continuous setting in the pioneering work of Palmov \cite{Palmov1967}, in the study of solidification. The discrete form presented above can be derived from the principle of virtual work. Since the incompatibilities are fixed, by \eqref{eq:1}, every virtual variation
\begin{equation}
\widetilde{\mathcal E}=\{(\widetilde\epsilon^i_x,\widetilde\epsilon^i_y),i=1,\ldots,n\}	
\end{equation}
of the strain state must satisfy
\begin{equation} 
\widetilde\epsilon_x^{i+1}-\widetilde\epsilon_x^i=0,
\end{equation}  
which imply that 
\begin{equation}
\widetilde\epsilon_x^i=\widetilde\epsilon_x,
\end{equation}
with $\widetilde\epsilon_x$ an arbitrary constant.
Thus, the internal virtual work associated to a virtual variation of the strains is
\begin{equation}
W[\widetilde{\mathcal E}]=	\sum_{i=1}^n\sigma^i_x\widetilde\epsilon^i_x+\sum_{i=1}^n\sigma^i_y\widetilde\epsilon^i_y=\Big(\sum_{i=1}^n\sigma^i_x\Big)\widetilde\epsilon+\sum_{i=1}^n\sigma^i_y\widetilde\epsilon^i_y.
\end{equation}
At equilibrium, in the absence of applied loads, the internal work must vanish for every virtual variation of the horizontal strains, whence $\eqref{2}$, since $\widetilde\epsilon_x$ and $\{\widetilde\epsilon_y^i,i=1,\ldots,n\}$ are arbitrary.

The linear system \eqref{1}--\eqref{2} is well-posed: it is comprised of linearly independent equations whose number equals that of the unknowns. The unique solution is:\begin{equation}
\label{3}
    \sigma_x^i = E
    \left(
    \sum_{k=1}^{i-1}\delta_k - \frac{1}{{n}}
    \sum_{a=1}^{n}\sum_{k=1}^{a-1}\delta_k
    \right)\qquad i=1,...,{n},
\end{equation}
where we use the convention that sums over an empty set of indices vanish, i.e., $\sum_{k=1}^{0}\delta_k=0$. The  solution \eqref{3} confirms that the quantities $\delta_i$ are source of stress, even in the absence of loading. This motivates our interpretation of $\delta_i$ as a lumped version of strain incompatibility.

\begin{figure}[h]
\centering
  \includegraphics[width=0.95\linewidth]{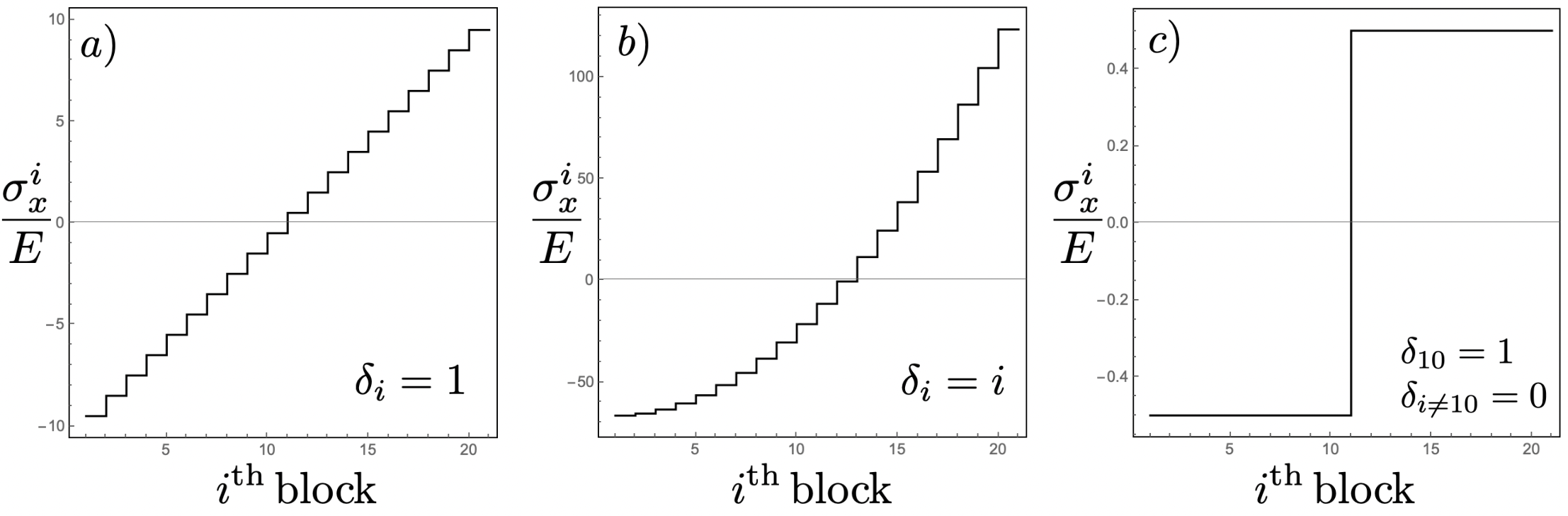} 
  \caption{\label{Fig3} \footnotesize{Renormalized (stress/Young modulus) residual stress in a stack with $n=20$ blocks. $a)$ constant incompatibilities $\delta_i=1$; $b)$ linear incompatibilities; $c)$ singular incompatibility.}}
\end{figure}

For illustrative purposes, sample distributions of horizontal stress obeying \eqref{4} for ${n}=20$ are represented in Fig.\ref{Fig3}. In the special situation when the incompatibility is constant ($\delta_i=\delta$), the stress is (see Fig.\ref{Fig3}$_a$): 
\begin{equation}
\label{4}
    \sigma_x^i = \left(i-\frac{{n}+1}{2}\right)E\delta \qquad i=1,...,{n}.
\end{equation}
Note that the blocks in the upper half are in tension along the horizontal direction; conversely, the blocks in the lower half of the stack are in compression.
Since equilibrium in the horizontal direction is imposed \emph{globally}, our model does not display a localized residual stress in the proximity of an isolated disarrangement, as one would expect.

To see this, consider for instance the stress profile in Fig.~\ref{Fig3}${}_{\rm c}$, which is obtained for a stack of $n=20$ blocks with $\delta_{10}=\delta>0$ and $\delta_i=0$. The horizontal stress is 
\begin{equation}
\label{4new}
    \sigma_x^i = 
    \left\{
    \begin{array}{lll}
    - E\delta/2 & i< 10\\
    E\delta/2 & i\geq 11. 
    \end{array}
    \right.
\end{equation}
The stress resultant in the horizontal direction would sum to zero, but the whole stack beneath and above the interface between the blocks $10$ and $11$ has uniform stress, whereas in a real system one would expect this stress to decay moving towards the boundaries. This limitation of the model could clearly be relieved in multiple ways at the discrete level, for example by incorporating concentrated shear springs between all blocks, but this would significantly complicate the model and therefore would affect its transparency.

%%{\color{gray}In the spirit of formulating the problem of \emph{designing the residual stress}, the system of ${n}$ equations in \eqref{3} can be used to formulate an \emph{inverse problem} where a list of \emph{target} horizontal stresses $\sigma_x^i$, $i=1,\ldots,{n}$ satisfying the first of \eqref{3} are \emph{given}, and the incompatibilities $\delta_i$ are the unknowns. Although there are ${n}$ equations, and only ${n}-1$ unknowns, the system \eqref{3} is not overdetermined, since \eqref{2} is exactly its solvability condition. Since the system is solvable, its solution is also unique. In particular, for null data, i.e., for vanishing target stresses, the unique solution is the trivial one, which implies that the only way to attain a null target stress is to have null incompatibility, as in the case of linear elasticity.\color{black}}\todo[inline]{Testo in grigio lo toglierei perche': 1) e' gia' stato detto; 2) il problema inverso viene introdotto piu' avanti.}

\subsection{Heavy blocks}

We now consider the case when, along with incompatibilities, an external loading is applied to the stack, which we ascribe to gravity. As a modelling assumption, we consider that at the interface between every pair of blocks there is a lumped  mass, as shown in Figure 4. 

The weights of the lumped masses, renormalized by the bock width, are denoted by the symbol $b_i$. These have the same dimensions of the 2D Young modulus $E$ and the 2D stress components $\sigma_{x/y}^i$, as a result of having dimensions of force per unit of length. As a remark, the $b_i$ should not be confused with 2D body forces (force/area), yet we continue to use the same symbol because of how these are physically related to weight. 

As depicted in Fig.\ref{Fig1}$_c$, heavy blocks made of a lumped mass topping a massless elastic block will be filled in gray, whereas massless blocks are unfilled. Note that as a further simplification in our model, thinking of the weight as concentrated supports the hypothesis that the heavy blocks undergo homogeneous strain. 
%usage of a lumped mass serves to justify the absence of inhomogeneous deformations inside heavy blocks.

Gravity is applied to the stack while the incompatibilities between the blocks are kept frozen to their prescribed values. Therefore, during the application of the loading, the blocks can not slide relative to each other, but they deform laterally as a whole. The equilibrium equations \eqref{3} are now replaced by:
\begin{equation}
\label{555}
\left\{
    \begin{array}{lll}
    {
    \displaystyle\sum_{i=1}^{n}\sigma_x^{i}=0},
    \\[1.5em]
    {
    \displaystyle\sigma_y^{i+1}-\sigma_y^{i}=b^i  \qquad i=1,...,n-1},\\[1em]
    {
    \displaystyle\sigma_y^{n} = -b^n}.
    \end{array}
\right.
\end{equation}

\begin{figure}[h]
\centering
  \includegraphics[width=0.8\linewidth]{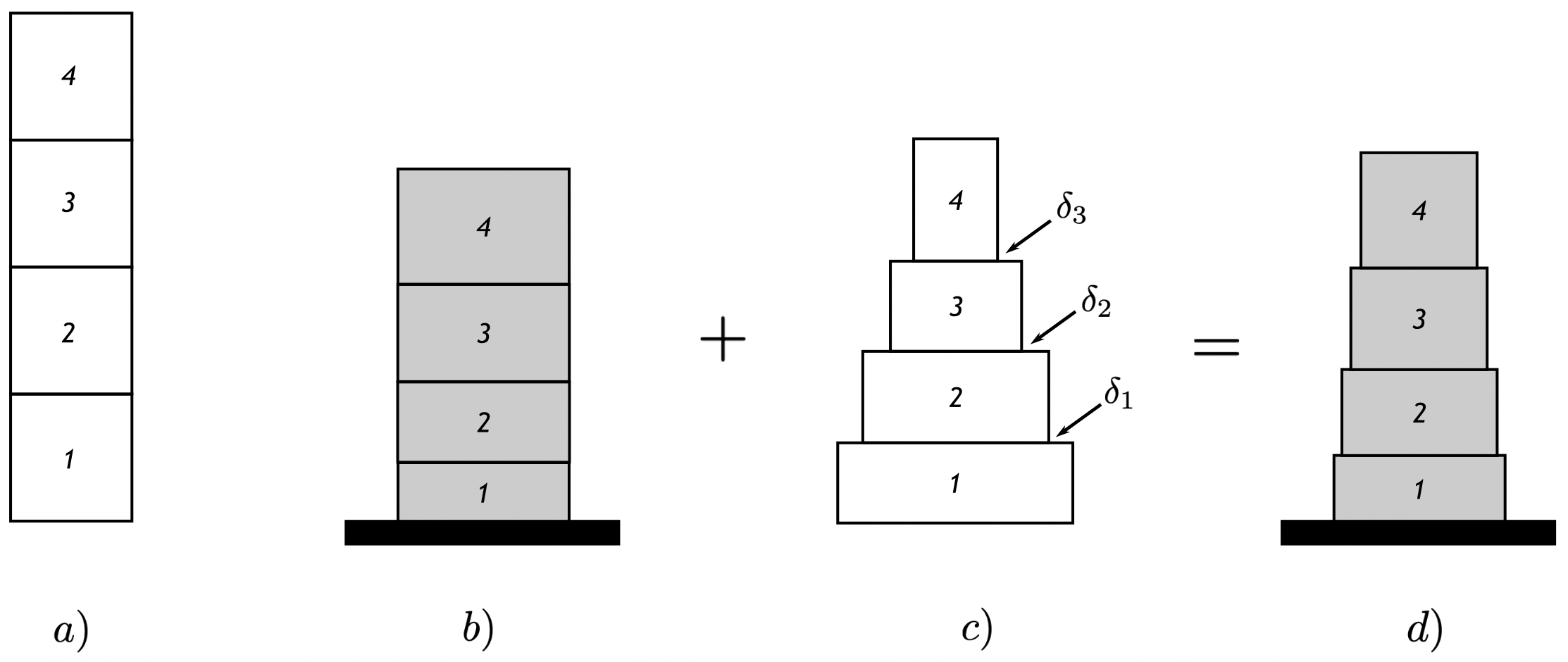}
\caption{\label{Fig4}a) \footnotesize{Equilibrium configuration of a stack with $b_i=0$ and $\delta_i=0$. b) Stack in equilibrium with $b_i>0$ and $\delta_i=0$. c) Stack in equilibrium with $b_i=0$ and $\delta_i\neq 0$. d) Combined effect of the weights and incompatibilities. In $b)$ and $c)$ the black line denote a smooth rigid foundation, that allows horizontal sliding but holds the structure vertically.}}
\end{figure} 

%In the system under analysis, the weights of the blocks $ b_i $ are applied by means of heavy plates to ensure that the blocks can deform homogeneously, as previously introduced, and there is no friction at the interface. With the blocks and plates stacked on top of each other, the system is statically determined along the vertical direction and solving the equilibrium problem along the $ y $ direction we obtain:
 The equilibrium equations determine the stresses along the vertical direction:
\begin{equation}
\label{6}
    \sigma_y^{i} =- \sum_{s=i}^n b_s  \qquad i=1,...,n. 
\end{equation}
The stresses in the horizontal direction are instead obtained by solving the first of \eqref{555}, together with the  constitutive equations \eqref{1} and the incompatibility constraints \eqref{eq:1}. The solution  is (the details of the derivation may be found in Appendix \ref{AppA}):
\begin{equation}
\label{778}
\sigma_x^i=\sum_{k=1}^{i-1}(E\delta_k+\nu b_k)-\frac{1}{{n}}\sum_{k=1}^{{n}-1}(({n}-k)(E\delta_k+\nu b_k)) \qquad i=1,...,{n}.
\end{equation}
The mechanical effect of the weights $b_i$ and of the incompatibilities $\delta_i$ is illustrated in the cartoon in Fig.\ref{Fig4}. Due to the linearity of the problem, the final shape represented in $d)$ may be seen as the result of superposing $a)$, wherein $\delta_i=0$ and $b_i\neq 0$, and $c)$, where instead $\delta_i\neq 0$ and $b_i= 0$.

It is worth noting that the weight $b_n$ of the last block \emph{does not affect the horizontal stress.} Similarly to what we did to obtain \eqref{4}, we can consider the special case when the incompatibilities and the weights are spatially uniform: $\delta_i=\delta$, $b_i=b$. In this case, the horizontal stress is an affine function of $i$:
\begin{equation}
\label{4bis}
    \sigma_x^i = \left(i-\frac{{n}+1}{2}\right)(E\delta+\nu b) \qquad i=1,...,{n},
\end{equation}
and the profile of the horizontal residual stress is qualitatively the same as in Figure 3. The horizontal strain is
\begin{equation}\label{eq:11}
\varepsilon_x^i=\left(i-\frac{n+1}2\right)\delta+\nu\frac{n+1}2\frac b E.
\end{equation}
We notice from \eqref{eq:11} that the effect of a constant weight is to shift the average value of the horizontal strain from zero. The scaling factor $(n+1)/2$ is due to the fact that the mechanical work performed along the loading path by the weight on top of the $i$-th block is increasing with $i$, due to a sort of ``telescopic effect''. Hence, the weights do not perform the same work.

A remarkable feature of the general solution \eqref{778} is that the dependence of the stress $\sigma_x^i$ on the incompatibilities $\delta_i$ and on the weights $b_i$ is functionally similar (apart from the different coefficients $E$ and $\nu$). Albeit this model is extremely simplified, it therefore retains a feature that is present at the level of the three-dimensional theory of elasticity, where incompatible strains and body forces play a similar role as sources of residual stress fields \cite{MarkenscoffGupta}.

Another interesting feature of the general solution \eqref{778} is that that horizontal stresses would vanish inside each block ($\sigma_x^i=0$) {\it if and only if}
\begin{equation}\label{deltai}
\delta_i = - \frac{\nu}{E} b_i. 
\end{equation}
Equation \eqref{deltai} also unveils that the incompatibilities $\delta_i$, if these can be externally controlled, could be used to anneal the horizontal stress in a heavy stack, through a sort of remodeling process.

If, instead, the incompatibilities can be nailed and maintained in the stack, one could exploit \eqref{deltai} to \emph{compensate} for the effect of the concentrated weights $b_i$, for example, to annihilate residual stresses in the deformed stack. Therefore, the loaded column would have zero horizontal stresses in all blocks, whereas of course the unloaded column (where weights are removed) would be residually stressed.

\section{Growth of an heavy stack: the direct problem\label{Sec3}}
While until now the incompatibilities were prescribed, in this section we establish how their distribution arises during a construction process of layered deposition, where a  hypothetical printing devices deposits the blocks one by one, through a sequence of steps, with the incompatibilities being locked at the end of each step by a layer of glue between adjacent blocks. We call this the \emph{direct problem}, as opposed to the {\it inverse problem} \cite{Zurlo2017,Truskinovsky2019}, which consists in determining the incompatibilities that are required to produce a desired distribution of residual stress.\

We describe the construction process incrementally. Making reference to the cartoon of Fig.\ref{Fig5}, the process proceeds as follows. At the first step of the construction process, the first block $j=1$ is laid on a frictionless support. For $b^1$ the weight of the block, the stress and strain state at the end of the first step are
\begin{equation}
\Sigma^1=\{(\sigma_x^{1,1},\sigma_y^{1,1})\}=\{(0,-b^1)\}.	
\end{equation}
The corresponding strains are
\begin{equation}
\mathcal E^1=\{(\epsilon_x^{1,1},\epsilon_y^{1,1})\},
\end{equation}
with
\begin{equation}
\epsilon_y^{1,1}=-\frac {b^1}E,\qquad \epsilon_x^{1,1}=-\nu \epsilon_y^{1,1}.
\end{equation}\color{black}
Now, suppose that we are at the end of the $j^{\text{th}}$ step (see Fig.~\ref{Fig5}.a). The stack is composed of $j$ blocks, the glue that bonds all the blocks is solidified, the incompatibilities are: 
$$
\Delta^{j}=(\delta_i, i=1,...,j-1). 
$$ 
Note that at the end of step $j=1$ there is no incompatibility.  The incompatibilities are \emph{locked}, and the stack is in equilibrium under the action of the weights $b_i$, $i=1,\ldots j$. In this configuration, equilibrium is described by a $j$-uple of horizontal and vertical stresses 
$$
\Sigma^j=\{(\sigma_x^{i,j},\sigma_y^{i,j}), i=1,...,j\}
$$ 
and by a $j$-uple of horizontal and vertical strains 
$$
\mathcal E^j=\{(\epsilon_x^{i,j},\epsilon_y^{i,j}),i=1,\ldots,j\}
$$ 
which solve the equilibrium problem discussed in the previous section, namely:
\begin{equation}
\label{99}
 (P_{j})\left\{
 \begin{array}{lll}
    {\color{black}
    \displaystyle\sum_{i=1}^{j}\sigma_x^{i,j}=0}\\[1.5em]
    {\color{black}
    \displaystyle\sigma_y^{i+1,j}-\sigma_y^{i,j}=b^i  \qquad i=1,...,j-1}\\[1em]
   {\color{black}
    \displaystyle\sigma_y^{j,j} = -b^{j}}\\[1em]
    {\color{black}
    \epsilon_{x/y}^{i,j}=\frac{\displaystyle\sigma_{x/y}^{i,j}-\nu\sigma_{y/x}^{i,j}}{\displaystyle E} \qquad i=1,...,j
    }\\[1em]
    {\color{black}
    \displaystyle\epsilon_x^{i+1,j}-\epsilon_x^{i,j}=\delta_i\qquad i=1,...,j-1}
    \end{array}
\right.
\end{equation}
The $(j+1)^{\rm th}$ construction step takes place in two stages. In the first stage, the printing 
device deposits the $(j+1)^{\text{th}}$ block on top of the stack, keeping the block 
under a self-balanced couple of ``deposition surface stress'':
\begin{equation}\label{21}
\mathring{\sigma}_x^{j+1}:=\sigma_x^{j+1,j}. 
\end{equation}
In addition, the printing device maintains the $(j+1)^{\text{th}}$ block in equilibrium, by supplying a vertical force
\begin{equation}
	\mathring\sigma_y^{j+1}=-b_{j+1}, 
\end{equation}
at the bottom face of the block, see Fig.\ref{Fig5}$_b$, which equilibrates the weight $b_{j+1}$.

  {In writing \eqref{21}, we take inspiration from previous work on continuous surface growth \cite{Zurlo2017,Truskinovsky2019,Zurlo2018}, where the surface (i.e. tangential) components of the stress are controlled at the growth surface, besides the usual tractions or displacements, ordinarily controlled in non-growing bodies. %When the printing For $b_{j+1}$ the weight of the $(j+1)^{\rm th}$ block, prior to deposition the printing device 
\begin{figure}[h]
\centering
  \includegraphics[width=0.75\linewidth]{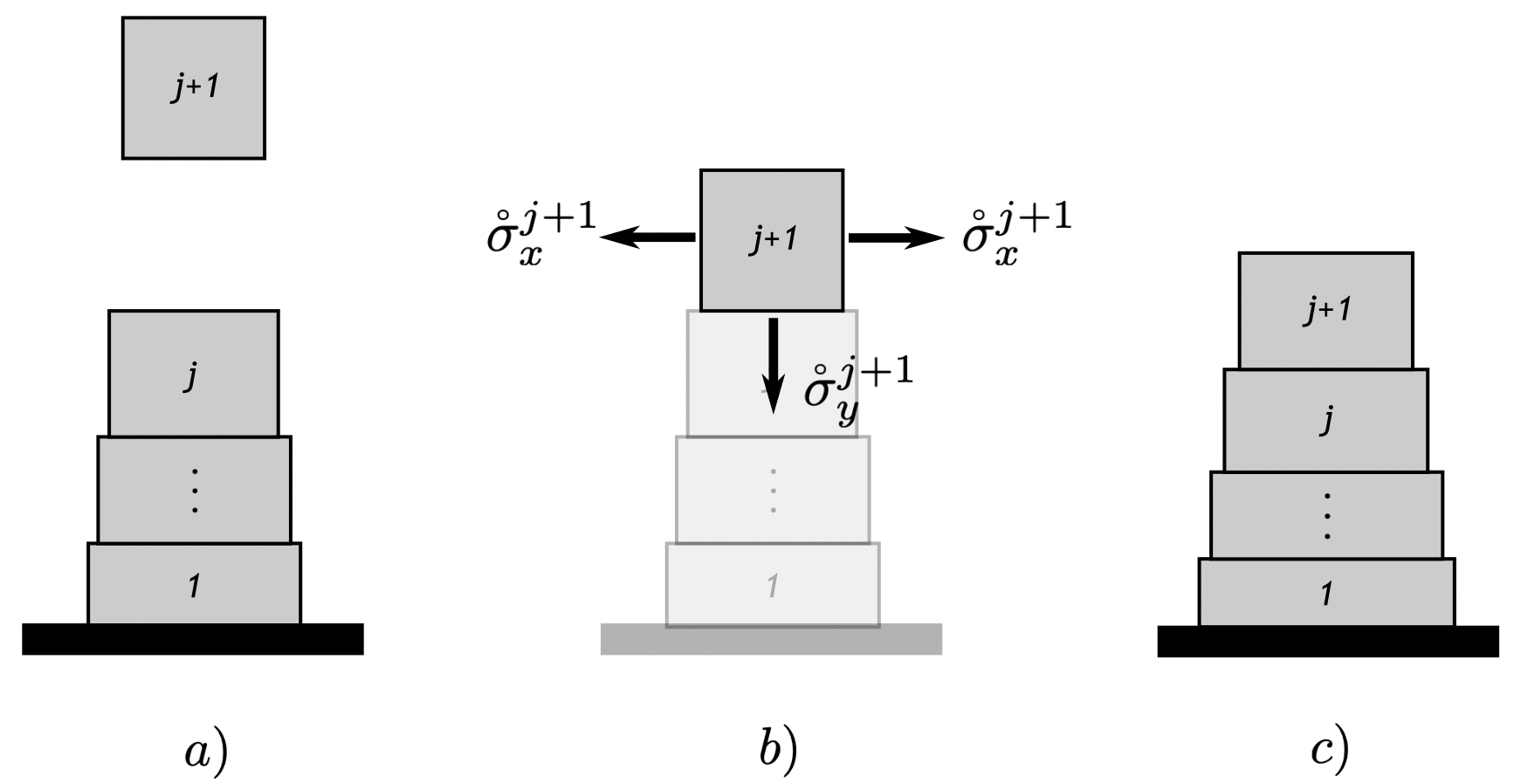} 
  \caption{\footnotesize{\label{Fig5}
 a) A block $j+1$ is prepared, in order to be deposited on top of an equilibrated stack of $j$ block. 
 b) Immediately prior to deposition, the ``printer'' pre-stresses the block through a self-balanced couple of horizontal forces of magnitude $\mathring\sigma_x^{j+1}$, and balances the weight of the block by exerting
a vertical force $\mathring\sigma_y^{j+1}$ on the the bottom of the block. 
 c) When the block is dropped, the new stack made of $j+1$ blocks reattains equilibrium. }}
\end{figure}
In this stage, the deposition strains of the upper block are: 
\begin{equation}\label{14}
\mathring\epsilon_x^{j+1}=\frac{\mathring\sigma_x^{j+1}+\nu b_{j+1}}{E},\qquad \mathring\epsilon_y\\
^{j+1}=-\frac{b_{j+1}+\nu\mathring\sigma_x^{j+1}}{E}.	
\end{equation}
We suppose that the glue begins its solidification process at the end of the first stage, as soon as the block $j+1$ is brought to contact with the stack.

In the second stage of step $j+1$, the printing devices releases the block, and the construction process pauses until the glue has solidified. At the end of the solidification process, the stack has attained a new equilibrium configuration (see Fig.\ref{Fig5}$_c$), described by a a new stress state
 \begin{equation}\Sigma^{j+1}=\{(\sigma_x^{i,j+1},\sigma_y^{i,j+1}), i=1,...,j+1\},\label{24}	
 \end{equation}
 and by a $j$-uple of horizontal and vertical strains 
\begin{equation}\label{25}
\mathcal E^{j+1}=\{(\epsilon_x^{i,j+1},\epsilon_y^{i,j+1}),i=1,\ldots,j+1\}.	
\end{equation}
As we shall see below, the new stress and strain distributions depend on the nature of the glue between block $j$ and block $j+1$. Whatever, these distributions be, one must expect that, in general, the stress $\sigma_x^{j+1,j+1}$ in the upper block after the attainment of equilibrium  be different from the pre-deposition stresses $\mathring{\sigma}_x^{j+1}$. This, as we will illustrate, has an important quantitative effect in the distribution of incompatibilities resulting from the deposition process.

The stress and strain states $\Sigma^{j+1}$ and $\mathcal E^{j+1}$, after the printing device is removed, and after the glue has solidified, depends on the solidification speed of the glue. We consider two extreme cases, which we refer to as ``fast glue'', which ``freezes'' the incompatibility at the very instant of deposition, and ``slow glue'', which allows the last block to slide on top of the second-to-last until the horizontal stress in the last block vanishes.

A point we would like to stress is that, even if at the current stage we have not yet described how the incompatibility between the last and second to last blocks is calculated, we will assume that incompatibilities between all pairs of block do not evolve during the process. In other words, their value is frozen at the value determined at deposition, and the latter will depend crucially on the mechanical behavior of the ``glue''.}

\subsection{Fast glue}
The ``fast glue'' solidifies at the precise instant when block $j+1$ is brought into contact with the stack, that is, when the printing device is still holding the block, i.e. the weight $b_{j+1}$ is not trasmitted to the stack yet. Prior to deposition, the equilibrium of the stack is not perturbed, hence the stresses and strains are those resulting from Problem $(P_j)$; moreover, the strain state of block $j+1$ is given by \eqref{14}. Thus, the incompatibility between block $j$ and block $j+1$ calculates as: 
\begin{equation}
\label{159}
\delta_{j}=\mathring\epsilon_x^{j+1}-\epsilon_x^{j,j}=-\frac{b_{j+1}+\nu \mathring{\sigma}_x^{j+1}}{E}-\epsilon_x^{j,j}.
\end{equation}
We remark that $\delta_j$ depends not only on the pre-stress $\mathring\sigma_x^{j+1}$ imparted on the new block by the printing device, but also, through Problem $(P_j)$, on the incompatibilities $(\delta_1,\ldots,\delta_{j-1})$, \emph{which keep memory of the history of the construction process}.

In Fig.\ref{Fig5} we can see the cartoon of the process. At deposition of the $(j+1)-$th block, when the latter is still supported by the printing machine, the fast glue perfects the bond, therefore locking the incompatibility at the value \eqref{159}. At this point, the printing device releases the block, and this in turn imparts both a vertical action and an horizontal action on the underlying stack, which at this instant is yet unbalanced. Thus, for the case of the fast glue, the new stress and strain distributions \eqref{24} and \eqref{25} are the solution of Problem $(P_{j+1})$, where the incompatibility distribution
\begin{equation}
\Delta^{j+1}=(\delta_1,\ldots,\delta_{j-1},\delta_j),	
\end{equation}
with $\delta_1,\ldots,\delta_n$ the same as in the previous step and $\delta_n$ given by \eqref{159}. The process is then iterated with the deposition of new blocks.

The direct problem, as formulated above, admits a closed-form solution. \color{black}The details of its derivation may be found in Appendix \ref{AppB}. In terms of deposition strains, the incompatibility $\delta_i$ is:
\begin{equation}
\label{17b}
\delta_i=\mathring\epsilon_x^{i+1}-\frac{i-1}{i}\mathring\epsilon_x^i-\frac{\nu b_i}{E}.
\end{equation}
Using \eqref{14}, we can rewrite \eqref{17b} as:
\begin{equation}
	\label{17}
	\delta_i=\frac{\mathring{\sigma}_x^{i+1}+\nu b_{i+1}}{E}-\frac{i-1}{i}\frac{\mathring{\sigma}_x^{i}+\nu b_{i}}{E}-\frac{\nu b_i}{E}.
\end{equation}
The finding that the incompatibility $\delta_i$ depends only on the last two elements $\mathring\sigma_x^i, \mathring\sigma_x^{i+1}$ of the deposition history $(\mathring\sigma_x^2,\ldots,\mathring\sigma_x^i, \mathring\sigma_x^{i+1})$, was to be expected. Indeed, even in the continuous case  the incompatibility depends on the spatial derivative (the divergence) of the pre-deposition surface stress \cite{Trincher,Zurlo2017,Zurlo2018}. Were this continuous problem to be discretized to find a numerical solution, the approximation of the divergence a the boundary would involve the values of the discretized deposition stress at the two grid points closest to the boundary, in accordance with \eqref{17}.

Once the incompatibilities are known, the horizontal stresses can be calculated from \eqref{778}, an equation that we repeat here for $n=j$:
%\begin{equation}
%\label{32}
%\sigma_x^{i,j+1}=\sum_{k=1}^{i-1}\left(E\delta_k +\nu b_k\right) -\frac{1}{j+1}\sum_{k=1}^{j}\left(\left(j+1-k\right)\left(E\delta_k+\nu b_k\right)\right),
%\end{equation}
\begin{equation}
\label{32}
\sigma_x^{i,j}=\sum_{k=1}^{i-1}\left(E\delta_k +\nu b_k\right) -\frac{1}{j}\sum_{k=1}^{j-1}\left(\left(j-k\right)\left(E\delta_k+\nu b_k\right)\right). 
\end{equation}

\bigskip

\subsection{Slow glue}
If the glue perfects the bond between the last and second to last blocks very slowly, we may think that the interface between the $j+1$-th and the $j$-th blocks is frictionless since the instant of deposition to the instant when the whole stack finally reaches equilibrium. Since the glue cannot support shear stresses prior to solidification, which occurs more slowly than the attainment of equilibrium,
the horizontal stress of block $j+1$ needs to vanish, 
\begin{equation}\label{sigma0slow}
\sigma_x^{j+1,j+1}=0. 
\end{equation}
The system of equations that govern the elastic state in the stack is
\begin{equation}
\label{18}
(\bar P_{j+1})\left\{
    \begin{array}{lll}
    \displaystyle\sum_{i=1}^{j}\sigma_x^{i,j+1}=0,\\[1em]
\displaystyle\sigma_y^{i+1,j}-\sigma_y^{i,j}=b^i,  \qquad i=1,...,j,\\[1em]
{\color{black}
\displaystyle\sigma_y^{j+1,j+1} = -b^{j+1}},\\[1em]
    \epsilon_{x/y}^{i,j+1}=\frac{(\sigma_{x/y}^{i,j+1}-\nu\sigma_{y/x}^{i,j+1})} {E}, \qquad i=1,...,j+1,
    \\[1em]
    {\color{black}
    \displaystyle\epsilon_x^{i+1,j+1}-\epsilon_x^{i,j+1}=\delta_i,\qquad i=1,...,j-1},\\[1em]
    \displaystyle\sigma_x^{j+1,j+1}=0.
    \end{array}
\right.
\end{equation}
When the glue perfects the bond at the attainment of equilibrium, the difference between the horizontal strains of the $j+1$-th and the $j$-th block remains frozen to its value at equilibrium, and the incompatibility $\delta_j$ is therefore set to:
\begin{equation}\label{deltajslow}
	\delta_j=\epsilon_x^{j+1,j+1}-\epsilon_x^{j,j+1}.
\end{equation}
Once the incompatibility is locked, the printing devices starts the deposition of the next block, and the procedure is iterated until $j\le n$.  

In summary, the process proceeds as follows (following again the cartoon of Fig.\ref{Fig5}): prior to deposition, the stack of $j$ blocks is in equilibrium, and the block $j+1$ is made available geometrically; the block is then prestretched and maintained in equilibrium by the printing device; at deposition, however, differently than the case of fast glue, the horizontal stress in the $j+1$ block drops to zero, and therefore the underlying stack deforms to reach a new equilibrium only due to the weight of the last block. Finally, the glue
dries and the incompatibility is frozen at the value defined by \eqref{deltajslow}. 

It is immediate to check (details in Appendix \ref{Appslow}) that the solution to the equilibrium condition \eqref{sigma0slow} leads to
\begin{equation}
\label{22}
\delta_i = -\frac{\nu}{E} b_i. 
\end{equation}
Equation \eqref{22}, which coincides with \eqref{deltai}, confirms that in the slow glue protocol, the incompatibility is determined by the local value of $b_i$. This was to be expected: the slow glue solidifies only \emph{after} equilibrium has been attained, meaning that prior to equilibrium the glue is frictionless, which is the same assumption at the basis of \eqref{deltai}, where the variables $\delta_i$ were permitted to evolve freely upon attainment of equilibrium. 

We conclude this section by observing that \eqref{778} with \eqref{22} imply that in the slow glue protocol,  the horizontal stresses $\sigma_x^{i,j}$ vanishes everywhere in the stack during the construction process. Note however that, since the incompatibilities $\delta_i$ remain frozen, the removal of the loads upon completion of the construction results, in general, into a non-trivial distribution of residual stress.\color{black}

\newpage

\section{Direct problem: examples\label{Sec4}}

To get further insight into the findings of the Sec.\ref{Sec3}, we are now going to consider some direct problems, where the final stress distributions is obtained under different protocols are deduced and compared quantitatively and qualitatively.

\subsection{Fast glue\label{Sec:fastglue}}

In the fast glue case, one can prescribe both the weight and the horizontal prestress of each new block. By using the solutions \eqref{17} and \eqref{32} it is now possible to compare the outcomes of different loading protocols. 

To establish a meaningful comparison, we consider stacks with the same final number $n$ of blocks and the same total weight. Specifically, we consider blocks with increasing, constant and decreasing weight, 
\begin{equation}
b_i = i E, \qquad
b_i = \frac{n+1}{2} E, \qquad
b_i = (n+1-i)E, 
\qquad (i=1,...,n). 
\end{equation}
Note that in all cases above, the total weight of the column is the same %$n(n+1)E/2$ 
but, as we shall see, the stress distributions are significantly different.

\begin{figure}[thb!]
\centering
\includegraphics[width=0.9\linewidth]{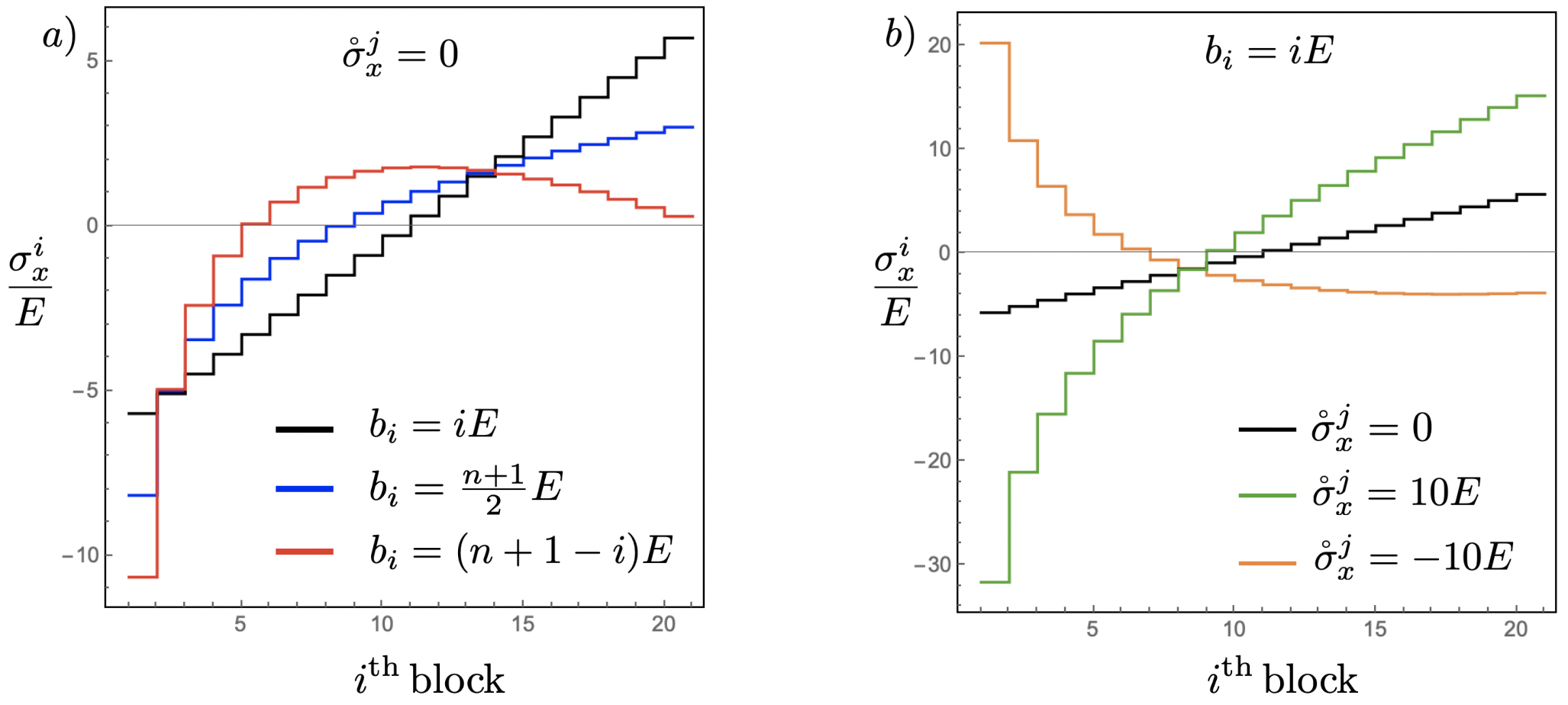} 
\caption{\label{Fig6} \footnotesize{Stress distributions resulting from the {\it fast glue} protocol in a stack with $n=300$ layers, with $\mathring\sigma_x^j=0$ and different distributions of $i$ sharing the same total weight (left); or with $b_i$ linear and different prescriptions for the deposition stress $\mathring\sigma_x^j$ (right).}}
\end{figure}

We first compare stacks that are manufactured with $\mathring{\sigma}_x^j=0$. The resulting stress distributions are reported in Fig.\ref{Fig6}. All the stress distributions have, of course, zero resultant. By changing the order of the weights one can change the qualitative behavior of the $\sigma_x^i$ in the stack. In particular, if the $b_i$ are linearly increasing the distribution of the $\sigma_x^i$ is also linearly increasing, if the $b_i$ are uniform the $\sigma_x^i$ still increase, but not linearly, and finally if the $b_i$ are decreasing,  the $\sigma_x^i$ are non-monotonic. 

Another interesting effect results from the comparison of stacks that are manufactured with the same linear distribution of $b_i= i\,E$, but that differ in terms of deposition stress $\mathring\sigma_x^j$. In this case one can obtain stacks where the resulting distribution of $\sigma_x^i$ is linear if $\mathring\sigma_x^j=0$, but it becomes concave or convex if, respectively, $\mathring\sigma_x^j>0$ or $\mathring\sigma_x^j<0$, that is, if the blocks are pre-tensioned or pre-compressed prior to deposition. 

In passing we note that in both all distributions considered above, there are special points where the various stress distributions intersect, two points in the case where $\mathring\sigma_x^j=0$ and the $b_i$ are changed, and one such point in the case where $b_i$ is linear and the $\mathring\sigma_x^j$ are changed.

\subsection{A remark on the (non) controllability of the stress in 3D printing}

A key aspect of the fast glue protocol is represented by the fact that the {\it pre-deposition} stress 
\begin{equation}
\mathring\sigma_x^{j}:=\sigma_x^{j,j-1}
\end{equation}
and the {\it post-deposition} stress $\sigma_x^{j,j}$ of the upper block will differ in general. This happens because after the deposition of the $j$-th block, the latter will deform together with the underlying stack to reach a new equilibrium configuration, and therefore its level of stress will generally change from the pre-deposition value. 
\begin{figure}[thb!]
\centering
\includegraphics[width=0.8\linewidth]{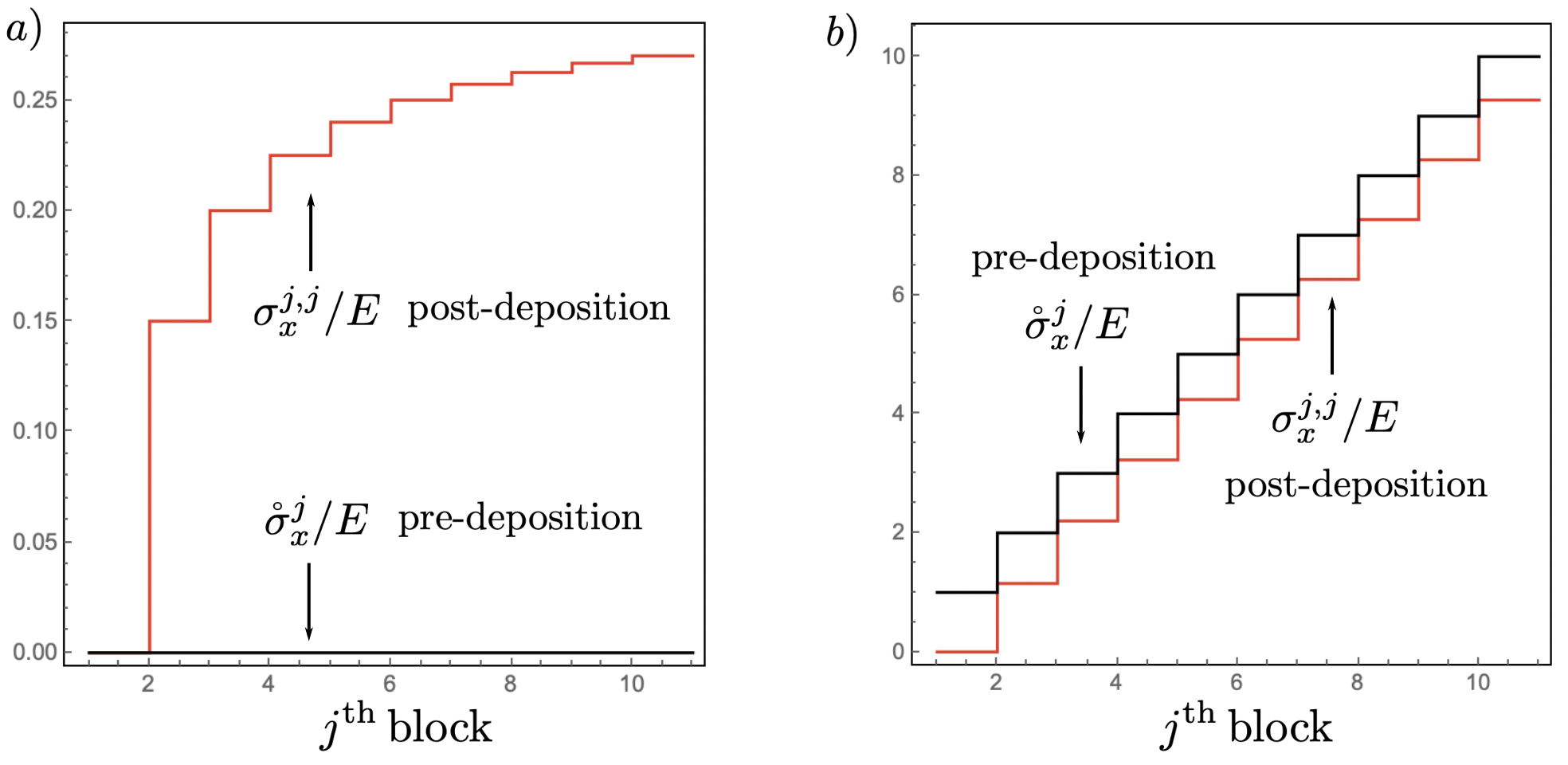} 
\caption{\label{Fig7} Differences between the {\it pre-deposition} stress $\mathring\sigma_x^{j}$ (black) and the resulting {\it post-deposition} stress $\sigma_x^{j,j}$ (red) in the newly deposited block, in two cases: $a)$: $\mathring\sigma_x^{j}=0$ and $b)$: $\mathring\sigma_x^{j}=j E$. In both cases the body force is uniform, $b_i=E$. The plots display the dimensionless ratio $\sigma_x/E$.}
\end{figure}

This feature is illustrated in Fig.\ref{Fig7} for two different loading scenarios, both taking place with uniform weights ($b_i=E$). In both cases the pre-deposition stress the post-deposition stress are different, although the differences seem to be bounded if the number of blocks is very large.  

The difference between the pre-deposition stress and the post-deposition stress has relevant implications on 3D printing. Indeed, this result shows that during additive manufacturing, the value of the surface stress right after deposition can not be controlled directly, since it will depend {\it in a non-local fashion} on the whole state of deformation of body. This fact was already been highlighted in pioneering works on surface growth, specifically we make reference to the seminal work of Trincher \cite{Trincher}. Further developments of this aspect were discussed in the setting of continuous incompatible surface growth in \cite{Zurlo2017}.

\subsection{Slow glue}

While in the fast glue protocol the pre-deposition stress and post-deposition stress can be both different than zero, and differ between each other, in the slow glue protocol the post-deposition stress $\sigma_x^{j,j}$ will always drop to zero throughout the process, despite any value given to the pre-deposition stress $\mathring\sigma_x^j$. Indeed, as Fig.\ref{Fig6}$_a$ shows, in the fast glue case even if the pre-deposition stress $\mathring\sigma_x^j$ is zero,  the post-deposition stress $\sigma_x^{j,j}$ will generally differ than zero, whereas it would always be zero in the slow glue case. 

\begin{figure}[thb!]
\centering
\includegraphics[width=0.9\linewidth]{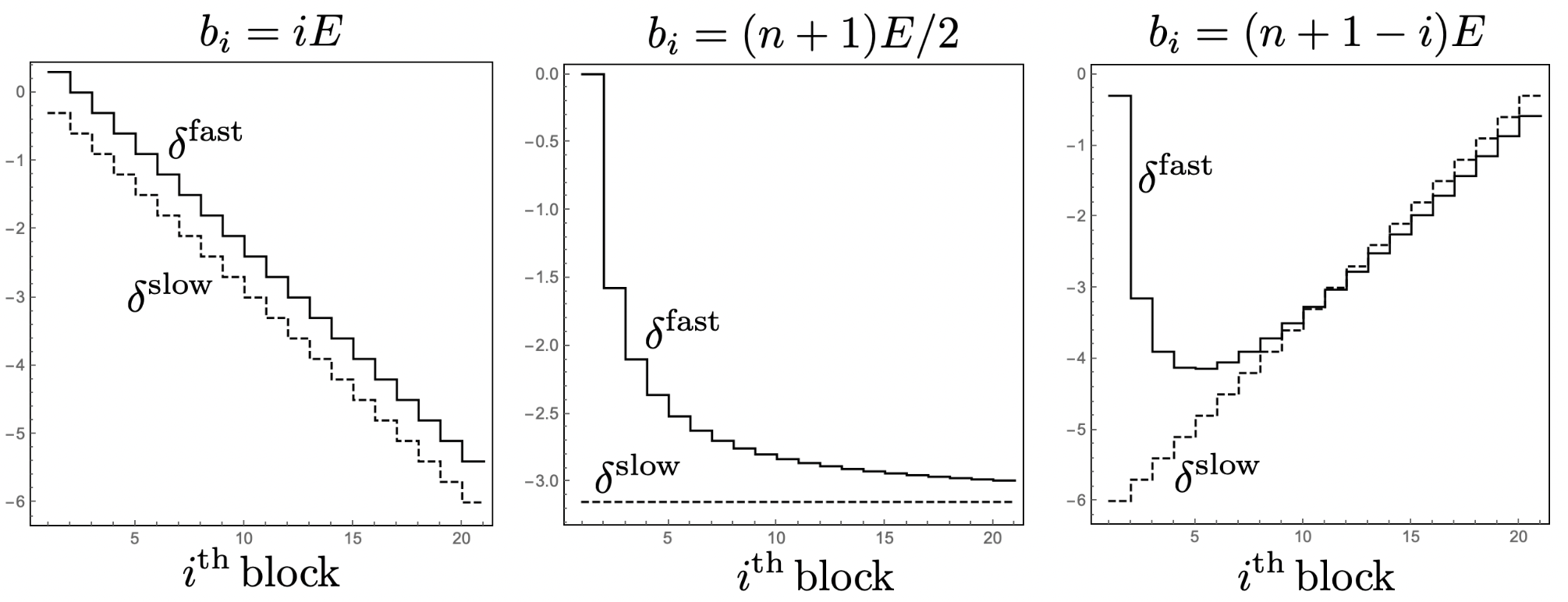} 
\caption{\label{Fig8} Comparison between the incompatibilities resulting from fast and slow protocols, for walls that are manufactured with $\mathring\sigma_x^j=0$. Here $\nu=0.3$.}
\end{figure}

\begin{figure}[thb!]
\centering
\includegraphics[width=0.95\linewidth]{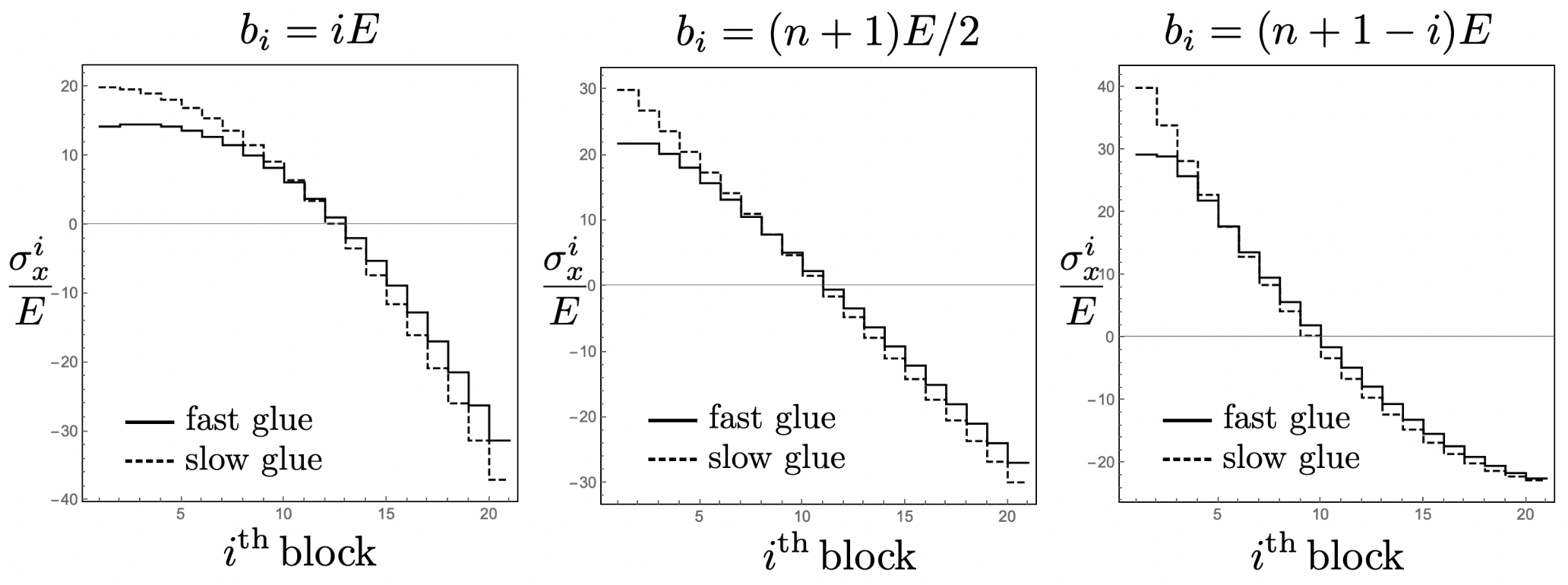} 
\caption{\label{Fig9} Comparison between the residual stresses arising from fast and slow protocols, for towers manufactured with $\mathring\sigma_x^j=0$. Here $\nu=0.3$.}
\end{figure}

This marks a strong difference between fast and slow glue protocols, that may be readily appreciated in 
by comparing the incompatibilities stored at the end of the process. Consider the accretion of a stack with $\mathring\sigma_x^j=0$ and, once again, three distributions of $b_i$ leading to the same total weight. As evidenced in Fig.\ref{Fig8}, in all cases the distributions of incompatibilities are different in the fast and slow glue protocols, above all for the first layers. Eventually however, for $b_i$ constant or decreasing, the incompatibilities would tend asymptotically to a common value if the number of layers is very large, whereas the difference in incompatibilities would remain constant in for $b_i$ linearly increasing.
Differences in terms of residual stress (that is, the stress distribution in the stack upon removal of all external loading) may be appreciated in all cases covered in Fig.\ref{Fig8} in the plots of Fig.\ref{Fig9}. Upon calculating the incompatibility, the stress distributions may be computed with $b_i=0$ from \eqref{778}. The results show that if the stacks that are manufactured with $\mathring\sigma_x^j=0$, the effects of fast/slow glue are non-negligible, but essentially of {\it quantitative} nature. 

\section{The inverse problem: programming residual stress}
Thus far we have explored the problem of finding the residual stress arising from a given deposition protocol. We now turn our attention to the following {\it inverse problem}: find the deposition protocol 
\begin{equation}
\Pi=(\mathring\sigma^2,\ldots,\mathring\sigma^n;b^1,\ldots,b^n)
\end{equation}
which delivers a desired residual stress state. To avoid confusing the target stress state with the current state of stress of the growing stack, we denote the former by
\begin{equation}\label{target}
T_x^n=(\tau_x^{1},\ldots,\tau_x^n),
\qquad\text{s.t.}\qquad
\sum_{i=1}^n\tau_x^i=0.
\end{equation} 
which is needed to achieve a desired residual stress state. 

We remark that the target state of stress is solely sourced by the incompatibilities, therefore if weights are applied on the stack during manufacturing, the attainment of the target state is achieved by suppressing all the weights. We also remark that by removal of the weights, the vertical stresses $\sigma_y^{i,n}$ 
vanish, therefore the target state is defined solely by a self-balanced distribution of horizontal stresses.

By using the definition \eqref{eq:1} and the constitutive equations \eqref{1} (or by solving \eqref{3}) we find that the incompatibilities needed to attain the target state are:
\begin{equation}
\label{20}
\delta_i=\frac{\tau_x^{i+1}-\tau_x^i}{E} \qquad i=1,...,n-1.
\end{equation}
Thus, our problem reduces to finding the deposition protocol that results in a given distribution of incompatibilities $(\delta_1,\ldots,\delta_{n-1})$. 

The protocol depends on whether we are using a fast glue or a slow glue, and on the loading applied to the stack during the deposition. {We remark that albeit the target state is unloaded, weights may still be used in the manufacturing process to produce incompatibilities, but of course such weights would be removed when the stack is complete. Weights are, of course, the only possible type of control in the slow glue case, since here the horizontal pre-deposition stress always drops to zero.}

\subsection{Fast glue: inverse problem}

{Consider the deposition of a heavy stack with the fast-glue protocol. By solving \eqref{17} where we now treat $\delta_i$ and $b_i$ as prescribed and $\mathring\sigma_x^i$ as unknown, and bearing in mind that $\mathring\sigma_x^1=0$, we obtain 
\begin{equation}\label{new}
\mathring\sigma_x^i=\frac{1}{i-1}\sum_{k=1}^{i-1}
\left(\nu\left(2 k b_k - b_k-b_{k+1}\right) + E k \delta_k\right), \qquad i>1. 
\end{equation}
The meaning of \eqref{new} is that to produce a desired distribution of incompatibility $\delta_i$ in a heavy stack with given weights $b_i$, one needs to provide the pre-deposition horizontal stress $\mathring\sigma_x^i$ as dictated by \eqref{new}. At the end of the process, the weights would be removed and the only source of stress in the stack are the incompatibilities. Note that if the stack is directly manufactured with $b_i=0$, the final distribution of stress would coincide precisely with the target residual stress state \eqref{target}.

As a first application, we manufacture a stack with sinusoidal distribution of residual stress, 
\begin{equation}\label{targettau}
\tau_x^i = \sin\left(\frac{2 \pi i}{n}\right)
\end{equation}
which is clearly self-balanced. The target residual stress is represented in Fig.\ref{Fig10}$_a$. Once the incompatibilities are calculated by \eqref{20}, by making use of \eqref{new} we can determine the required deposition protocol. 

For illustration, we show that the same target may be achieved by two different processes, involving both heavy and massless stacks. In particular we manufacture the stack: 
\begin{itemize}
\item[1)] through the deposition of pre-stretched massless layers; 
\item[2)] through the deposition of pre-stretched heavy layers with weight increasing as $b_i=i E$. 
\end{itemize}
\begin{figure}[thb!]
\centering
\includegraphics[width=0.9\linewidth]{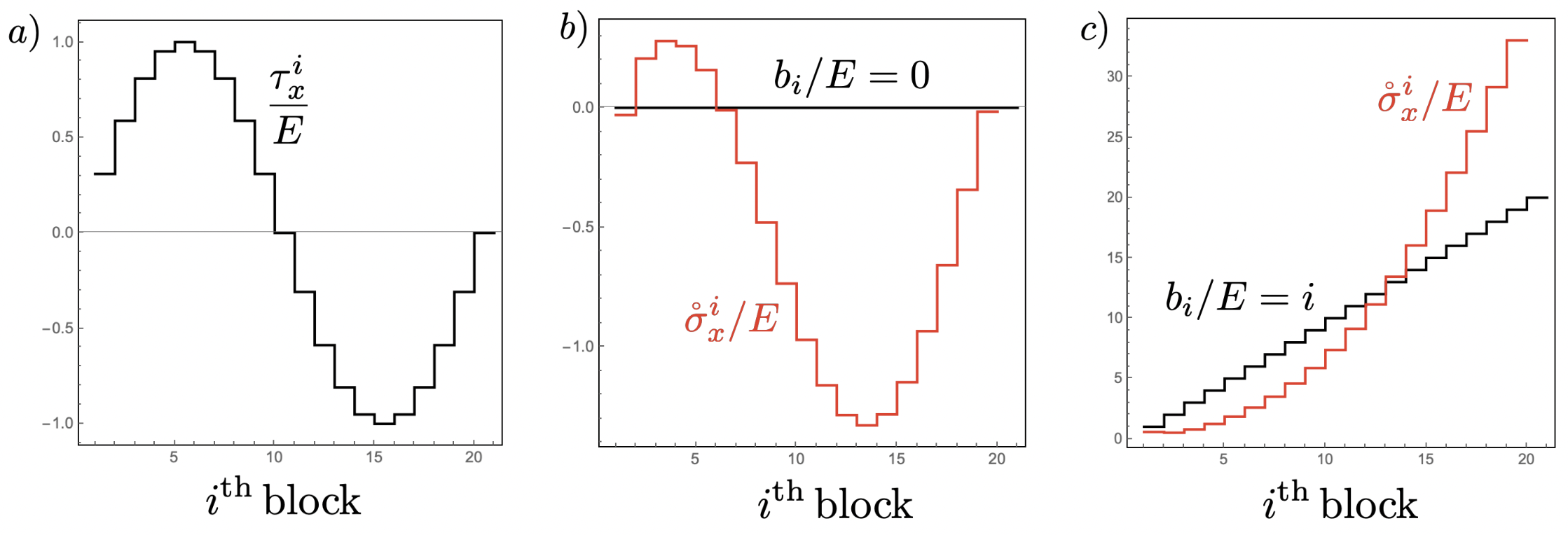} 
\caption{\label{Fig10} $a)$ target residual stress $\tau_x^i$ to be achieved through the fast glue protocol. $b)$ Required deposition pre-stress to achieve the target in the case of massless weights, and $c)$ required deposition pre-stress in the case of linearly increasing weights. Here $\nu=0.3$.}
\end{figure}

The outcomes are illustrated in \ref{Fig10} for a stack of $n=20$ blocks. In Fig.\ref{Fig10}$_{b}$ we illustrate the case $1)$, in Fig.\ref{Fig10}$_{c}$ we illustrate case $2)$. The required deposition stress seems intuitively consistent with the final stress distribution in the case $b_i=0$, but it is rather different from the target stress in the case $b_i=i E$. 

An interesting remark relative to \eqref{new} is that the distribution of deposition stresses $\mathring\sigma_x^i$ required to produce a given incompatibility is independent on the weights, if these  satisfy 
\begin{equation}
b_k+b_{k+1}=2kb_k,
\end{equation}
giving
\begin{equation}
\tilde{b}_i =b_1 \frac{2^{i-1}}{\Gamma(i)}{\prod_{k=0}^{i-2}(\tfrac{1}{2}+k)},
\end{equation}
where $\Gamma(i)$ is Euler's Gamma function, with $b_1$ arbitrary. For $b_1=0$ we obtain trivially the case of weightless blocks, but it is interesting to note that even for $b_1\neq 0$ the deposition stress required to produce a certain target residual stress would not depend on the weight distribution. 
}

\subsection{Slow glue: inverse problem}

{In the slow glue paradigm the condition $\sigma_x(i,i)=0$ results into the constraint $E\delta_i+\nu b_i=0$ between weights and incompatibilities, therefore during deposition that weights may be used to obtain a target residual stress $\tau_x^i$. Indeed, we simply need to prescribe weights according to 
\begin{equation}\label{bslow}
b_i=\frac{\tau_x^i-\tau_x^{i+1}}{\nu}. 
\end{equation}
This makes the treatment of the inverse problem for the slow-glue case definitely simpler than the fast-glue case. Moreover, at least conceptually, \eqref{bslow} shows that any distribution of $\tau_i$ can be targeted through the deposition of weights $b_i$. 

There is, however, a subtlety in the practical implementation of this method. Indeed, since $\tau_x^i$ must have zero average, it means that $\tau_x^i$ cannot be nether constant, nor strictly increasing or decreasing on the stack. This implies that there will be zones, where $\tau_x^i$ is increasing, where one would need to deposit blocks with a negative weight.}

\section{Conclusions}

In this work we have formulated a discrete model that captures, in a minimal setting where the material is treated as homogeneous, linearly elastic and isotropic, and where plasticity and thermal effects are neglected, some fundamental features of additively manufactured solids. By introducing a simplified kinematics that allows for the onset of incompatible deformations between the layers (schematized as elastic blocks) of an additively manufactured stack, we have studied the role of deposition prestress applied to the blocks, and of the weight of the blocks, on the resulting state of incompatibility. From here, we have computed the final residual stress patterns, and we could single out the factors behind qualitative and quantitative differences emerging in their distributions. 

To capture with minimal complexity the role played an ``ideal glue'' between the blocks - which is required to maintain incompatible deformations, therefore preventing the blocks to slide relative to each other  - we have devised two extreme behaviors: one, called ``fast glue'', where the bond between the prestressed block and the underlying stack occurs while the new block is still maintained in equilibrium by some invisible ``printing device''; the other, called ``slow glue'', where the block is first released and is left free to find an equilibrium configuration together with the underlying stack, before the glue perfects the bond. 

The model proposed in this work is a very elementary approximation of the complex aspects that occur in the process of additive manufacturing, for example the role of the glue is overly simplified and the deformations are constrained in a very special way, however its analytical simplicity allows to obtain closed form results, that are very transparent in physical terms. In particular, the model retains many features of richer continuous models, like in particular the presence of a new type of boundary condition, that is the control of ``surface stress'', that is not contemplated in non-growing solids \cite{Zurlo2017,Zurlo2018,Truskinovsky2019}. 

The main findings of the model have been illustrated by comparing effects due to different deposition protocols, showing that some controls (for example the type of glue) will play a quantitative effect in the final stress distribution of the stack, whereas other controls (for example the deposition surface pre-stress) will play a qualitative effect. These results pave the way for new studies on the fundamental role of the adhesive properties of the growth surface. 

The last part of the manuscript is devoted to the formulation of a technologically important inverse problem, consisting in finding the  deposition protocol that is required to achieve a desired distribution of residual stress, but also heterogeneous elastic properties, in an additively manufactured stack  \cite{Yan2017,Agnelli2020,Danescu2013}. 

The proposed model should be conceived as a template, that has the advantage to highlight with physical and analytical transparency the non-trivial mechanics involved in additive manufacturing of prestressed solids.

\newpage

\color{black}

\section{Appendix}
\subsection{Derivation of  \eqref{778}\label{AppA}}
This section illustrates the steps to arrive at the determination of the relation \eqref{778}. Starting from the first relation of system \eqref{555}, together with the constitutive equations \eqref{1} and \eqref{eq:1} and taking into account that the problem along the vertical direction is solved by: $ \sigma_y ^{i}=- \sum_{s=i}^{n} b_s$. The problem along the horizontal direction can be expressed by the following system:

  \begin{equation}
\label{313}
\left\{
    \begin{array}{lll}
    {
    \displaystyle\sum_{i=1}^{n}\sigma_x^{i}=0}\\
     \\
    {
    \epsilon_{x/y}^{i}=\frac{(\sigma_{x/y}^{i}-\nu\sigma_{y/x}^{i})}{E} \qquad i=1,...,n
    }\\
    \\
    {
    \displaystyle\epsilon_x^{i+1}-\epsilon_x^{i}=\delta_i\qquad i=1,...,n-1}\\
  
    \end{array}
\right.
\end{equation}
Expressing the third relation of the system \eqref{313} through the constitutive relation we can write:
\begin{equation}
\label{314}
\sigma_x^{i+1}-\sigma_x^{i}=E\delta_i+\nu b_i
\end{equation}
From which we can write:
\begin{equation}
\label{315}
\sigma_x^{i+1}=\sigma_x^{1}+\sum_{k=1}^{i}E\delta_k+\sum_{k=1}^{i}\nu b_k
\end{equation}
Using now the first relation of the system \eqref{313} we determine the following relation:
\begin{equation}
\label{316}
\sigma_x^{1}=-\frac{1}{n}\sum_{i=2}^{n}\sum_{k=1}^{i-1}E\delta_k-\frac{1}{n}\sum_{i=2}^{n}\sum_{k=1}^{i-1}\nu b_k
\end{equation}
Replacing the \eqref{316} in \eqref{315} we get:
\begin{equation}
\label{317}
\sigma_x^{i+1}=\sum_{k=1}^{i}E\delta_k-\frac{1}{n}\sum_{a=2}^{n}\sum_{k=1}^{a-1}E\delta_k+\sum_{k=1}^{i}\nu b_k-\frac{1}{n}\sum_{a=2}^{n}\sum_{k=1}^{a-1}\nu b_k,
\end{equation}
from which we can write the following relation:
\begin{equation}
\label{318}
\sigma_x^i=E\left(\sum_{k=1}^{i-1}\delta_k-\frac{1}{{n}}\sum_{a=2}^{{n}}\sum_{k=1}^{a-1}\delta_k\right)+\nu\left(\sum_{k=1}^{i-1}b_k-\frac{1}{{n}}\sum_{a=2}^{{n}}\sum_{k=1}^{a-1}b_k\right) 
\end{equation}
rearranged it we can write:
\begin{equation}
\label{319}
\sigma_x^i=\sum_{k=1}^{i-1}(E\delta_k+\nu b_k)-\frac{1}{{n}}\sum_{k=1}^{{n}-1}(({n}-k)(E\delta_k+\nu b_k)) 
\end{equation}

\subsection{Derivation of \eqref{17}\label{AppB}}
This section illustrates the steps to arrive at the determination of the relation \eqref{17}. Starting from the problem $P_j$, the solution of the problem along the vertical direction is that expressed by $\sigma_y^{i,j+1}=-\sum_{k=i}^{j+1}b_k$. The problem along the horizontal direction can be expressed by the following system:
 
  \begin{equation}
\label{303}
\left\{
    \begin{array}{lll}
    {
    \displaystyle\sum_{i=1}^{j}\sigma_x^{i,j}=0}\\
     \\
    {
    \epsilon_{x/y}^{i,j}=\frac{(\sigma_{x/y}^{i,j}-\nu\sigma_{y/x}^{i,j})}{E} \qquad i=1,...,j
    }\\
    \\
    {
    \displaystyle\epsilon_x^{i+1,j}-\epsilon_x^{i,j}=\delta_i\qquad i=1,...,j-1}\\
  
    \end{array}
\right.
\end{equation}
Expressing the third relation of the system \eqref{303} through the constitutive relation we can write:
\begin{equation}
\label{304}
\sigma_x^{i+1,j}-\sigma_x^{i,j}=E\delta_i+\nu b_i
\end{equation}
From which we can write:
\begin{equation}
\label{305}
\sigma_x^{i+1,j}=\sigma_x^{1,j}+\sum_{k=1}^{i}E\delta_k+\sum_{k=1}^{i}\nu b_k
\end{equation}
Using now the first relation of the system \eqref{303} we determine the following relation:
\begin{equation}
\label{306}
\sigma_x^{1,j}=-\frac{1}{j}\sum_{i=2}^{j}\sum_{k=1}^{i-1}E\delta_k-\frac{1}{j}\sum_{i=2}^{j}\sum_{k=1}^{i-1}\nu b_k
\end{equation}
Replacing the \eqref{306} in \eqref{305} we get:
\begin{equation}
\label{307}
\sigma_x^{i+1,j}=\sum_{k=1}^{i}E\delta_k-\frac{1}{j}\sum_{l=2}^{j}\sum_{k=1}^{l-1}E\delta_k+\sum_{k=1}^{i}\nu b_k-\frac{1}{j}\sum_{l=2}^{j}\sum_{k=1}^{l-1}\nu b_k
\end{equation}
Using the constitutive relation we arrive at the relation 
\begin{equation}
\label{308}
\epsilon_x^{j,j}=\frac{1}{E}(\sigma_x^{j,j}-\nu\sigma_y^{j,j})=\sum_{k=1}^{j-1}\delta_k-\frac{1}{j}\sum_{l=2}^{j}\sum_{k=1}^{l-1}\delta_k+\frac{1}{E}\sum_{k=1}^{j-1}\nu b_k-\frac{1}{jE}\sum_{l=2}^{j}\sum_{k=1}^{l-1}\nu b_k+\frac{\nu b_j}{E}
\end{equation}
which describes the horizontal deformation of the $j-th$ block when in the tower there are $j$ blocks.
From \eqref{308} we come to the incompatibility $ \delta_j$:
\begin{equation}
\label{309}
\delta_j=\mathring\epsilon_x^{j+1} -\sum_{k=1}^{j-1}\delta_k +\frac{1}{j}\sum_{k=1}^{j-1}(j-k)\delta_k-\frac{1}{E}\sum_{k=1}^{j-1}\nu b_k+\frac{\nu}{jE}\sum_{k=1}^{j-1}(j-k)b_k-\frac{\nu b_j}{E}
\end{equation}
The \eqref{309} can be rewritten as:
\begin{equation}
\label{310}
\delta_j=\mathring\epsilon_x^{j+1}-\frac{1}{j}\sum_{k=1}^{j-1}k\delta_k-\frac{\nu}{jE}\sum_{k=1}^{j-1}kb_k-\frac{\nu b_j}{E}
\end{equation}
We can further rewrite the formulation by writing:
\begin{equation}
\label{311}
\begin{split}
\delta_j&=\mathring\epsilon_x^{j+1}-\frac{1}{j}(j-1)\delta_{j-1}-\frac{1}{j}\sum_{k=1}^{j-2}k\delta_k-\frac{\nu}{jE}\sum_{k=1}^{j-1}kb_k-\frac{\nu b_j}{E}\\
&=\mathring\epsilon_x^{j+1}-\frac{j-1}{j}(\mathring\epsilon_x^j-\frac{1}{j-1}\sum_{k=1}^{j-2}k\delta_k-\frac{\nu}{(j-1)E}\sum_{k=1}^{j-2}kb_k-\frac{\nu b_{j-1}}{E})-\frac{1}{j}\sum_{k=1}^{j-2}k\delta_k-\frac{\nu}{jE}\sum_{k=1}^{j-1}kb_k-\frac{\nu b_j}{E}\\
&=\mathring\epsilon_x^{j+1}-\frac{j-1}{j}\mathring\epsilon_x^j+\frac{1}{j}\sum_{k=1}^{j-2}k\delta_k+\frac{\nu}{jE}\sum_{k=1}^{j-2}kb_k+\frac {j-1}{j}\frac{\nu b_{j-1}}{E}-\frac{1}{j}\sum_{k=1}^{j-2}k\delta_k-\frac{\nu}{jE}\sum_{k=1}^{j-1}kb_k-\frac{\nu b_j}{E}\\
&=\mathring\epsilon_x^{j+1}-\frac{j-1}{j}\mathring\epsilon_x^j+\frac{\nu}{jE}\sum_{k=1}^{j-2}kb_k+\frac {j-1}{j}\frac{\nu b_{j-1}}{E}-\frac{\nu}{jE}\sum_{k=1}^{j-1}kb_k-\frac{\nu b_j}{E}\\
&=\mathring\epsilon_x^{j+1}-\frac{j-1}{j}\mathring\epsilon_x^j+\frac {j-1}{j}\frac{\nu b_{j-1}}{E}-\frac{j-1}j\frac{\nu b_{j-1}}{E}-\frac{\nu b_j}{E}
\end{split}
\end{equation}
Simplifying the \eqref{311} we arrive at 
\begin{equation}
\label{312}
\delta_j=\mathring\epsilon_x^{j+1}-\frac{j-1}{j}\mathring\epsilon_x^j-\frac{\nu b_j}{E},
\end{equation}
a relation which expresses the incompatibility in the case of fast glue.

\subsection{Derivation of \eqref{22}\label{Appslow}}

To prove that in the slow glue protocol the incompatibility arising in the process is determined only by the weight of the underlying block, we recall that in the slow glue case the horizontal stress of the last block has to be zero at equilibrium. By imposing $\sigma_x^{i,i}=0$ for all $i\geq 2$, we obtain the system 
\begin{equation}
\begin{array}{lll}
\sigma_x^{1,1} = 0\\
\sigma_x^{2,2} = (E\delta_1+\nu b_1) = 0 \\
\sigma_x^{3,3} = (E\delta_1+\nu b_1) + 2(E\delta_2+\nu b_2) = 0 \\
\sigma_x^{4,4} = (E\delta_1+\nu b_1) + 2(E\delta_2+\nu b_2) + 3(E\delta_3+\nu b_3) = 0 \\
\dots
\end{array}
\end{equation}
Therefore, the result $E\delta_i+\nu b_i=0$ for all $i\geq 1$ follows at once. 

\color{black}

\section*{Acknowledgements}

SM thanks Regione Lazio for funding the project H-S3D (DTC 2021-2023 - TE1 Centro di Eccellenza, CUP F85F21001090003) and European Union for funding the project Cultural Heritage Active Innovation for Next-Gen Sustainable Society - CHANGES - PE00000020 PNRR – NextGenerationEU (CUP: F83C22001650006).

DR and SM thanks Regione Lazio for funding the projects: 3DH-solutions (Progetti di Gruppi di Ricerca 2020, CUP F85F21001530009). 

GZ gratefully acknowledges the support of GNFM (Gruppo Nazionale di Fisica Matematica) of the INdAM F. Severi. 

%%%%%%%%%%%%%%%%%%%%%%
%%%%%%%%%%%%%%%%%%%%%%


\begin{thebibliography}{100}

\bibitem{Palmov1967}
V. A. Pal'mov,
\textit{On stresses originating during material solidification},
\textit{Inzh. Zh. Mekh. Tverd. Tela}, \textbf{4}, 80--85 (1967).

\bibitem{Trincher}
V. K. Trincher,
\textit{Formulation of the problem of determining the stress-strain state of a growing body},
\textit{Izv. AN SSSR. Mekhanika Tverdogo Tela}, \textbf{19}(2), 119--124 (1984).

\bibitem{MarkenscoffGupta}
Xanthippi Markenscoff and Anurag Gupta,
\textit{Configurational balance laws for incompatibility in stress space},
\textit{Proc. R. Soc. A}, \textbf{463}, 1379--1392 (2007).

\bibitem{Abeyaratne2022}
R. Abeyaratne, E. Puntel, and G. Tomassetti,
\textit{Steady accretion of an elastic body on a hard spherical surface and the notion of a four-dimensional reference space},
\textit{Journal of the Mechanics and Physics of Solids}, \textbf{167}, 104958 (2022).

\bibitem{Abeyaratne2022a}
R. Abeyaratne, E. Puntel, and G. Tomassetti,
\textit{Surface accretion of a pre-stretched half-space: Biot's problem revisited},
\textit{J. Mech. Phys. Solids}, \textbf{167}, 104958 (2022).

\bibitem{Abeyaratne2022b}
R. Abeyaratne, E. Puntel, and G. Tomassetti,
\textit{On the Stability of Surface Growth: The Effect of a Compliant Surrounding Medium},
\textit{J. Elast.}, Preprint (2022).

\bibitem{Ackerly1989}
S. C. Ackerly,
\textit{Kinematics of accretionary shell growth, with examples from brachiopods and molluscs},
\textit{Paleobiology}, \textbf{15}(2), 147-164 (1989).

\bibitem{Agnelli2020}
F. Agnelli, A. Constantinescu, and G. Nika,
\textit{Design and testing of 3D-printed micro-architectured polymer materials exhibiting a negative Poisson's ratio},
\textit{Continuum Mechanics and Thermodynamics}, \textbf{32}(2), 433-449 (2020).

\bibitem{Ambrosi2007}
D. Ambrosi and F. Guana,
\textit{Stress-Modulated Growth},
\textit{Mathematical Mechanics of Solids}, \textbf{12}(3), 319-342 (2007).

\bibitem{Amar1986}
M. B. Amar and Y. Pomeau,
\textit{Theory of dendritic growth in a weakly undercooled melt},
\textit{Europhysics Letters}, \textbf{2}(4), 307 (1986).

\bibitem{Bacigalupo2012}
A. Bacigalupo and L. Gambarotta,
\textit{Effects of layered accretion on the mechanics of masonry structures},
\textit{Mechanics Based Design of Structures and Machines}, \textbf{40}(2), 163-184 (2012).

\bibitem{Brown1963}
C. B. Brown and L. E. Goodman,
\textit{Gravitational stresses in accreted bodies},
\textit{Proceedings of the Royal Society of London. Series A. Mathematical and Physical Sciences}, \textbf{276}(1367), 571-576 (1963).

\bibitem{Ciarletta2013}
P. Ciarletta, L. Preziosi, and G. A. Maugin,
\textit{Mechanobiology of interfacial growth},
\textit{Journal of the Mechanics and Physics of Solids}, \textbf{61}(3), 852-872 (2013).

\bibitem{Cowin2004}
S. C. Cowin,
\textit{Tissue growth and remodeling},
\textit{Annual Review of Biomedical Engineering}, \textbf{6}(1), 77-107 (2004).

\bibitem{darwinHorizontalThrustMass1883}
G. H. Darwin,
\textit{On the Horizontal Thrust of a Mass of Sand},
\textit{Minutes of the Proceedings of the Institution of Civil Engineers}, \textbf{71}(1883), 350-378 (1883).

\bibitem{Dafalias2008}
Yannis F. Dafalias, Dimitrios E. Panayotounakos, and Zacharias Pitouras,
\textit{Stress field due to elastic mass growth on spherical and cylindrical substrates},
\textit{Int. J. Solids Struct.}, \textbf{45}(17), 4629-4647 (2008).

\bibitem{Dafalias2009}
Yannis F. Dafalias and Zacharias Pitouras,
\textit{Stress field in actin gel growing on spherical substrate},
\textit{Biomech. Model. Mechanobiol.}, \textbf{8}(1), 9-24 (2009).

\bibitem{Danescu2013}
A. Danescu, C. Chevalier, G. Grenet, P. Regreny, X. Letartre, and J. L. Leclercq,
\textit{Spherical curves design for micro-origami using intrinsic stress relaxation},
\textit{Applied Physics Letters}, \textbf{102}(12), 123111 (2013).

\bibitem{DiCarlo2002}
A. Di Carlo and S. Quiligotti,
\textit{Growth and balance},
\textit{Mechanics Research Communications}, \textbf{29}(6), 449-456 (2002).

\bibitem{Epstein2000}
M. Epstein and G. A. Maugin,
\textit{Thermomechanics of volumetric growth in uniform bodies},
\textit{International Journal of Plasticity}, \textbf{16}(7-8), 951-978 (2000).

\bibitem{Garikipati2004}
K. Garikipati, E. M. Arruda, K. Grosh, H. Narayanan, S. Calve, S. Socrate, G. S. Kassab, and J. A. Weiss,
\textit{A continuum treatment of growth in biological tissue: The coupling of mass transport and mechanics},
\textit{Journal of the Mechanics and Physics of Solids}, \textbf{52}(7), 1595-1625 (2004).

\bibitem{Ge2016}
Q. Ge, A. H. Sakhaei, H. Lee, C. K. Dunn, N. X. Fang, and M. L. Dunn,
\textit{Multimaterial 4D printing with tailorable shape memory polymers},
\textit{Scientific Reports}, \textbf{6}(1), 1-11 (2016).

\bibitem{Humphrey2003}
J. D. Humphrey,
\textit{Continuum biomechanics of soft biological tissues},
\textit{Proceedings of the Royal Society A: Mathematical, Physical and Engineering Sciences}, \textbf{459}(2029), 3-46 (2003).

\bibitem{Kessler1988}
D. A. Kessler, J. Koplik, and H. Levine,
\textit{Pattern selection in fingered growth phenomena},
\textit{Advances in Physics}, \textbf{37}(3), 255-339 (1988).

\bibitem{Langer1980}
J. S. Langer,
\textit{Instabilities and pattern formation in crystal growth},
\textit{Reviews of Modern Physics}, \textbf{52}(1), 1-28 (1980).

\bibitem{Moulton2012}
D. E. Moulton, A. Goriely, and R. Chirat,
\textit{Mechanical growth and morphogenesis of seashells},
\textit{Journal of Theoretical Biology}, \textbf{311}, 69-79 (2012).

\bibitem{Moulton2014}
D. E. Moulton and A. Goriely,
\textit{Surface growth kinematics via local curve evolution},
\textit{Journal of Mathematical Biology}, \textbf{68}(1-2), 81-108 (2014).

\bibitem{Rashba1953}
E. I. Rashba,
\textit{Stress determination in bulks due to own weight taking into account the construction sequence},
\textit{Proc. Inst. Struct. Mech. Acad. Sci. Ukrainian SSR}, \textbf{18}, 23-27 (1953).

\bibitem{Reinhardt1892}
M. O. Reinhardt,
\textit{Das Wachsthum der Pilzhyphen},
\textit{Jahrbücher für wissenschaftliche Botanik}, \textbf{23}, 38-49 (1892).

\bibitem{Rodriguez1994}
E. K. Rodriguez, A. Hoger, and A. D. McCulloch,
\textit{Stress-dependent finite growth in soft elastic tissues},
\textit{Journal of Biomechanics}, \textbf{27}(4), 455-467 (1994).

\bibitem{Skalak1997}
R. Skalak, L. Zhao, and H. Wu,
\textit{Kinematics of surface growth},
\textit{Journal of Mathematical Biology}, \textbf{35}(8), 869-907 (1997).

\bibitem{Taber1995}
L. A. Taber,
\textit{Biomechanics of growth, remodeling and morphogenesis},
\textit{Applied Mechanics Reviews}, \textbf{48}(10), 487-545 (1995).

\bibitem{Tomassetti2016}
G. Tomassetti, T. Cohen, and R. Abeyaratne,
\textit{Steady accretion of an elastic body on a hard spherical surface and the notion of a four-dimensional reference space},
\textit{Journal of the Mechanics and Physics of Solids}, \textbf{96}, 333-352 (2016).

\bibitem{Truskinovsky2019}
L. Truskinovsky and G. Zurlo,
\textit{Nonlinear elasticity of incompatible surface growth},
\textit{Physical Review E}, \textbf{99}(5), 053001 (2019).

\bibitem{Mitchell2016}
M. R. Mitchell, T. Tlusty, and S. Leibler,
\textit{Strain analysis of protein structures and low dimensionality of mechanical allosteric couplings},
\textit{Proceedings of the National Academy of Sciences of the United States of America}, \textbf{113}(40), E5847-E5855 (2016).

\bibitem{Sozio2020}
F. Sozio, M. F. Shojaei, S. Sadik, and A. Yavari,
\textit{Nonlinear mechanics of thermoelastic accretion},
\textit{Zeitschrift für Angewandte Mathematik und Physik (ZAMP)}, \textbf{71}(3), 87 (2020).

\bibitem{Sozio2019}
F. Sozio and A. Yavari,
\textit{Nonlinear mechanics of accretion},
\textit{Journal of Nonlinear Science}, \textbf{29}(4), 1813-1863 (2019).

\bibitem{Sozio2017}
F. Sozio and A. Yavari,
\textit{Nonlinear mechanics of surface growth for cylindrical and spherical elastic bodies},
\textit{Journal of the Mechanics and Physics of Solids}, \textbf{98}, 12-48 (2017).

\bibitem{Yan2017}
L. Yan, M. Zhu, Y. Liu, Y. Huang, T. R. Nayak, Y. Hong, D. C. Patel, J. Huang, S. Chiba, N. X. Fang, and X. Zhang,
\textit{Architecture and coevolution of allosteric materials},
\textit{Proceedings of the National Academy of Sciences of the United States of America}, \textbf{114}(10), 2526-2531 (2017).

\bibitem{Zaza2021}
D. Zaza, M. Ciavarella, and G. Zurlo,
\textit{Strain incompatibility as a source of residual stress in welding and additive manufacturing},
\textit{European Journal of Mechanics - A/Solids}, \textbf{85}, 104147 (2021).

\bibitem{Zurlo2017}
G. Zurlo and L. Truskinovsky,
\textit{Printing non-Euclidean solids},
\textit{Physical Review Letters}, \textbf{119}(4), 048001 (2017).

\bibitem{Zurlo2018}
G. Zurlo and L. Truskinovsky,
\textit{Inelastic surface growth},
\textit{Mechanics Research Communications}, \textbf{93}, 174-179 (2018).

\end{thebibliography}
\end{document}